\documentclass[sigplan,nonacm]{acmart}
\renewcommand\footnotetextcopyrightpermission[1]{}

\AtBeginDocument{%
  }

\setcopyright{acmlicensed}
\copyrightyear{2026}
\acmYear{2026}
\acmDOI{XXXXXXX.XXXXXXX}

\acmSubmissionID{123-A56-BU3}

\usepackage{graphicx}
\usepackage{subcaption}
\usepackage{booktabs}
\usepackage{tabularx}
\usepackage{array}
\usepackage{ragged2e}
\usepackage{enumitem}
\usepackage{algorithm}
\usepackage{algpseudocode}
\usepackage{hyperref}
\hypersetup{
    colorlinks=true,
    linkcolor=blue,
    citecolor=blue
}
\usepackage{comment}

\usepackage{placeins}
\usepackage{float}

\graphicspath{{./figures/}}
\newcolumntype{L}{>{\RaggedRight\arraybackslash}X}
\newcolumntype{P}[1]{>{\RaggedRight\arraybackslash}p{#1}}

\usepackage{CJKutf8}

\usepackage{xcolor}

\newcommand{\sys}{HyperParallel-MoE}

\newif\ifshowcomment
\showcommenttrue

\ifshowcomment
  \newcommand{\cheng}[1]{\textcolor{red}{[\begin{CJK}{UTF8}{gbsn}Cheng:~#1\end{CJK}]}}
  \newcommand{\zw}[1]{\textcolor{blue}{[\begin{CJK}{UTF8}{gbsn}ZW:~#1\end{CJK}]}}
  \newcommand{\su}[1]{\textcolor{green}{[\begin{CJK}{UTF8}{gbsn}Su:~#1\end{CJK}]}}
  \newcommand{\hr}[1]{\textcolor{purple}{[\begin{CJK}{UTF8}{gbsn}Haoran:~#1\end{CJK}]}}
  \newcommand{\gp}[1]{\textcolor{magenta}{[\begin{CJK}{UTF8}{gbsn}GP:~#1\end{CJK}]}}
  \newcommand{\ck}[1]{\textcolor{orange}{[\begin{CJK}{UTF8}{gbsn}CK:~#1\end{CJK}]}}
\else
  \newcommand{\cheng}[1]{}
  \newcommand{\zw}[1]{}
  \newcommand{\su}[1]{}
  \newcommand{\hr}[1]{}
  \newcommand{\gp}[1]{}
  \newcommand{\ck}[1]{}
\fi

\begin{document}

\title{\sys: Multi-Core Interleaved Scheduling for Fast MoE Training on Ascend NPUs}

\newif\ifrelease
\newcommand{\para}[1]{\noindent {\bf #1} \hspace{2pt}}

\newcommand{\affmark}[1]{\textsuperscript{\ensuremath{#1}}}
\newcommand{\equalmark}{\textsuperscript{*}}
\newcommand{\internmark}{\textsuperscript{\ensuremath{\dagger}}}

\author{Zewen Jin\affmark{\clubsuit}\equalmark, Congkun Ai\affmark{\clubsuit}\equalmark, Guangpeng Zhang\affmark{\heartsuit}\equalmark, Hanbo Zhang\affmark{\diamondsuit}\internmark, Haoran Wang\affmark{\heartsuit}, \\Shihan Xiao\affmark{\heartsuit}, Da Lei\affmark{\heartsuit}, Xuefeng Jin\affmark{\heartsuit}, Teng Su\affmark{\heartsuit}, Cheng Li\affmark{\clubsuit}\affmark{\spadesuit}\\
\affmark{\clubsuit} University of Science and Technology of China \quad \\
\affmark{\heartsuit} Huawei Technologies Co., Ltd \quad \\
\affmark{\diamondsuit} Peking University \quad \\
\affmark{\spadesuit} Institute of Artificial Intelligence, Hefei Comprehensive National Science Center}
\renewcommand{\shortauthors}{Jin et al.}
\begin{abstract}


Modern Mixture-of-Experts (MoE) models increasingly rely on large-scale AI accelerator clusters for efficient training. Ascend NPUs expose heterogeneous on-chip compute resources, including matrix-oriented AIC units and vector-oriented AIV units with explicit cross-queue synchronization support. However, existing training frameworks largely execute MoE operators in a serialized kernel-by-kernel manner, leaving substantial heterogeneous parallelism underutilized.

This paper presents \sys, a compilation and scheduling framework for MoE training on Ascend NPUs. \sys\ transforms operator-level MoE execution into a statically scheduled tile-level heterogeneous taskflow spanning AIC and AIV resources. It introduces AIV-driven one-sided communication to eliminate host-side collective synchronization, dependency-preserving tile task generation to unify communication and computation under a common task abstraction, and event-driven static scheduling to coordinate cross-queue execution with low runtime overhead. \sys\ further executes the compiled taskflow within a unified runtime that concurrently drives AIC and AIV workers inside a single kernel launch, enabling fine-grained overlap among communication, matrix computation, and vector computation while preserving existing optimized operators. We implement \sys\ in the MindSpore and MindFormers stack and evaluate it using DeepSeek-style MoE models on Ascend A3 clusters. Across multiple expert-parallel configurations, \sys\ reduces Dispatch-to-Combine MoE-FFN latency by up to \(1.58\times\), demonstrating that tile-level heterogeneous scheduling can substantially improve MoE training efficiency on modern NPUs. The source code is available at \url{https://gitcode.com/mindspore/hyper-parallel/tree/master/hyper_parallel/core/multicore}.

\end{abstract}

\settopmatter{printfolios=true}
\maketitle
\begingroup
\renewcommand{\thefootnote}{*}\footnotetext{Equal contribution.}
\renewcommand{\thefootnote}{\ensuremath{\dagger}}\footnotetext{Work done during an internship at Huawei.}
\endgroup
\pagestyle{plain}

\section{Introduction}
\label{sect:intro}

Mixture-of-Experts (MoE) architectures have become a dominant design choice for modern large language models (LLMs), including DeepSeek-V3, Mixtral, and Qwen2.5-MoE~\cite{outrageous,switch,deepseekmoe,deepseekv2,deepseekv3,mixtral,qwen25}, because they scale model capacity without proportionally increasing per-token computation. As model sizes continue to grow toward hundreds of billions or even trillions of parameters, efficient MoE training increasingly relies on large clusters of AI accelerators~\cite{megatronlm,megascale}. At such scale, training efficiency is no longer determined solely by raw compute throughput, but also by how effectively frameworks exploit heterogeneous on-chip resources, overlap communication with computation, and reduce synchronization overheads. Consequently, recent system efforts have begun exploring fine-grained scheduling and communication-computation overlap to improve utilization of modern AI accelerators~\cite{pipedream,pipethreader,flashmoe,tritondistributed,flux,comet,lancet,deepep}.

Ascend NPUs are an increasingly adopted class of AI accelerators that expose a heterogeneous on-chip execution model~\cite{enec,squeezing,accelerating,deepserve}. Each Ascend AI Core contains physically decoupled AI Cube (AIC) and AI Vector (AIV) execution units: AIC specializes in matrix-intensive operators such as GEMM, while AIV executes vector computation, data movement, and communication-related operations~\cite{squeezing,accelerating}. Unlike conventional homogeneous execution models, Ascend NPUs provide independent hardware queues for Cube and Vector tasks, together with explicit event-based synchronization primitives. This architecture enables tile-level heterogeneous scheduling, where matrix computation, vector computation, and communication tasks can execute concurrently on different resources within the same kernel execution window. In addition, AIC and AIV share a large on-chip L2 cache, creating opportunities for cross-unit intermediate-result reuse without returning data to HBM.

However, existing training frameworks~\cite{megatronlm,mindspore} remain largely unaware of these architectural features, leading to suboptimal hardware utilization. Current frameworks typically run MoE-FFN as a kernel-by-kernel pipeline: Dispatch, GMM, SwiGLU, and Combine are each launched as separate kernels. AIC and AIV units therefore often idle in alternation rather than execute concurrently. Our profiling of a DeepSeek-V3-style model~\cite{deepseekv3} on 256 Ascend A3 NPUs reveals several inefficiencies. Although GMM operators dominate execution time, AIC units achieve only about 67\% MAC utilization on average. Meanwhile, Vector-side operators on the critical path still account for roughly 18\% of end-to-end training time, while EP communication contributes an additional 17\% overhead. More importantly, only about 61\% of EP communication latency can be hidden behind Cube computation, leaving the remaining 39\% exposed on the critical path. These inefficiencies stem from the mismatch between existing framework scheduling models and Ascend’s heterogeneous execution architecture.

In this paper, we present \sys, a compilation and scheduling framework for MoE training on Ascend NPUs. \sys\ transforms the originally operator-serialized MoE-FFN computation graph into a statically scheduled tile-level heterogeneous taskflow spanning AIC and AIV resources. Rather than isolating communication and computation behind global barriers, \sys\ decomposes both into fine-grained tile tasks and runs them concurrently within a unified execution window. Achieving this goal requires addressing three major challenges. First, conventional collectives impose global barriers and block fine-grained overlap between communication and computation. Second, tile decomposition must preserve both operator-level optimization constraints and cross-operator dependency correctness. Third, runtime scheduling must simultaneously satisfy correctness, low overhead, and high-quality heterogeneous overlap. Existing GPU-oriented fusion approaches either assume homogeneous SM architectures or require intrusive rewrites of production-grade operators, making them unsuitable for Ascend’s heterogeneous AIC/AIV execution model~\cite{flashmoe,tritondistributed,flux,comet,lancet,deepep,pipethreader,uniep,megamoe}.

To address the communication challenge, \sys\ introduces AIV-driven one-sided communication. Rather than relying on host-orchestrated collective communication, \sys\ converts EP AllToAll into device-side schedulable communication tasks driven directly by AIV workers. Each communication tile independently performs remote memory writes and updates completion signals through lightweight event counters, enabling communication progression to overlap naturally with computation at tile granularity. As a result, downstream GMM tasks can begin execution as soon as their required communication tiles become available, without waiting for the completion of the entire collective communication phase.

To address task decomposition and scheduling challenges, \sys\ introduces efficient dependency-aware tile task generation and static heterogeneous scheduling. \sys\ first constructs tile tasks from an Operator Dependency Graph (ODG), ensuring that communication, matrix computation, and vector computation share aligned tile boundaries and explicit dependency relationships. The framework then compiles these tasks into statically scheduled Cube Task Queue (CTQ) and Vector Task Queue (VTQ) taskflows augmented with event-driven synchronization semantics. Runtime execution therefore reduces to lightweight queue consumption and event waiting, while all dependency analysis, task generation, and execution ordering decisions are resolved offline. We integrate \sys\ into the MindSpore and MindFormers training stack~\cite{mindspore,mindformers} with low code intrusion, while preserving existing optimized implementations of GMM, SwiGLU, and communication operators.

We evaluate \sys\ using DeepSeek-V3-style MoE models~\cite{deepseekv3} on clusters of Ascend A3 NPUs. Across EP4, EP8, and EP16 settings, \sys\ reduces Dispatch-to-Combine MoE-FFN latency by \(1.49\times\)–\(1.58\times\) under balanced routing and achieves \(1.08\times\)–\(1.09\times\) speedup in end-to-end training under sampled natural routing. These results demonstrate that exposing heterogeneous AIC/AIV concurrency through tile-level static scheduling can substantially improve MoE training efficiency on existing Ascend hardware.

In summary, this paper makes the following contributions:

\begin{itemize}
    \item We propose a tile-level heterogeneous task abstraction for MoE training on Ascend NPUs, enabling communication, matrix computation, and vector computation to be jointly represented and scheduled within a unified taskflow.
    
    \item We design a static scheduling framework with event-driven synchronization that compiles MoE-FFN execution into precomputed CTQ/VTQ taskflows, enabling low-overhead heterogeneous execution while preserving dependency correctness.
    
    \item We implement \sys\ in the training stack of MindSpore and MindFormers  and demonstrate its effectiveness and generality on DeepSeek-style MoE models, achieving substantial MoE-FFN speedups on Ascend A3 clusters.
\end{itemize}


\section{Background and Motivation}
\label{sect:background}


\begin{figure}[t]
  \centering
  \includegraphics[width=\linewidth]{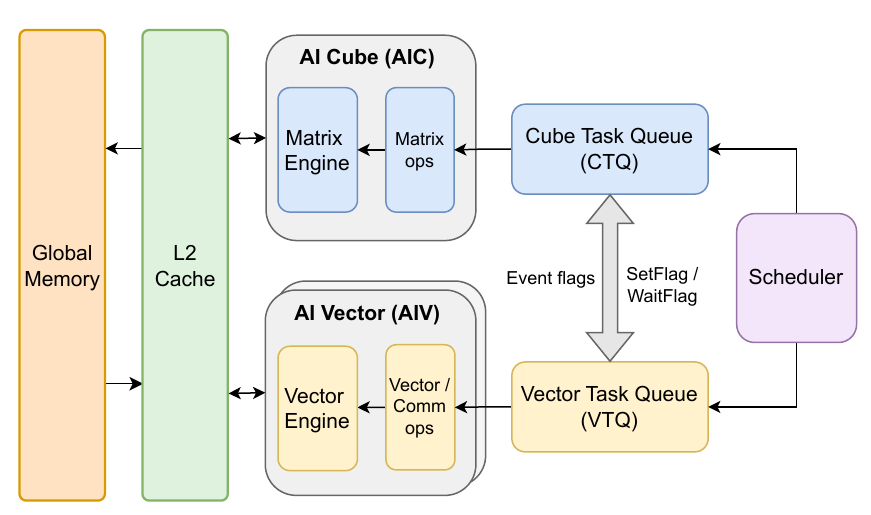}
  \caption{Ascend NPU heterogeneous AIC/AIV execution model.}
  \label{fig:arch}
\end{figure}

\subsection{Ascend NPU Architecture}
\label{sec:arch}

\noindent Ascend NPUs are advanced processors for neural network
computations that have seen increasing adoption~\cite{enec,squeezing,accelerating,deepserve}. As shown in Figure~\ref{fig:arch}, they feature a heterogeneous 
multi-core architecture, where each AI Core contains one AI Cube (AIC) unit and two 
AI Vector (AIV) units. AIC units are optimized for matrix operations, including GEMM 
and batched matrix multiplication, whereas AIV units execute vector-style operators, 
such as data movement and communication signaling. As an example, the recent 
Ascend A3 NPU consists of 25 AI Cores, providing a total of 25 AIC units and 50 AIV 
units. 


The AIC and AIV units are physically independent and are driven by separate instruction queues: the Cube Task Queue (CTQ) for AIC and the Vector Task Queue (VTQ) for AIV. Ascend NPUs further provide a programmer-controlled dispatch interface for heterogeneous scheduling~\cite{squeezing,accelerating}. First, AIC and AIV units concurrently fetch and execute tile-level tasks from their respective queues, enabling computations from both resource types to proceed in parallel whenever no data dependencies exist. Second, the architecture allows precise cross-queue dependency enforcement through event synchronization primitives, i.e., \texttt{CrossCoreSetFlag} and \texttt{CrossCoreWaitFlag} establish fine-grained dependencies between CTQ and VTQ. This lets matrix processing and vector processing form a tile-level pipeline within a single kernel. Overall, the explicit programming interface enables developers to directly specify, at tile granularity, which unit type handles which computation, precisely directing tasks into the dedicated hardware queues and realizing a fine-grained, event-driven scheduling paradigm that orchestrates on-chip heterogeneous resources with tight control.

Beyond synchronization via signals, the two types of cores share an on-chip L2 cache (192 MB) that is substantially larger than the data footprint of a single tile~\cite{squeezing,accelerating}. Consequently, when an AIC unit finishes a matrix multiply tile and writes the output to the L2 cache, a subsequent AIV vector tile can directly read that data from the L2, eliminating the need to write back to or read from the global
High Bandwidth Memory (HBM). Note that the L2 cache delivers more than 4$\times$ the read 
bandwidth of HBM. This on-chip data path, in which matrix multiply output lands in L2 and is then consumed by vector operations, turns the L2 cache into a high-bandwidth shared buffer that bridges the AIC and AIV pipelines, serving as a key foundation for efficient on-chip heterogeneous concurrency.


\begin{figure}[!t]
  \centering
  \includegraphics[width=0.95\linewidth]{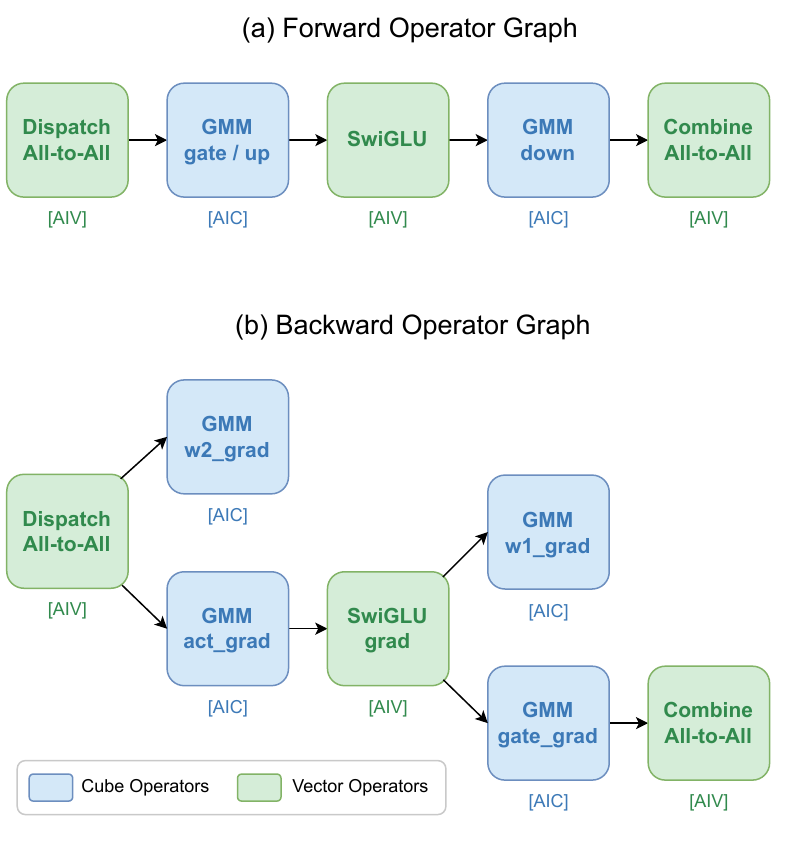}
  \caption{Forward and backward MoE-FFN operator graph with AIC/AIV mapping.}
  \label{fig:dag}
\end{figure}

\subsection{Mixture-of-Experts (MoE) Models}
\label{sec:moe}


\noindent
Modern large language models increasingly adopt Mixture-of-Experts (MoE) architectures~\cite{outrageous, switch} to scale model capacity without proportionally increasing the computation performed per token. An MoE layer contains a router and a set of expert networks. For each input token, the router selects a small subset of experts, typically through top-$k$ routing; only the selected experts process the token, and their outputs are weighted by routing scores and combined to form the final MoE output. Representative MoE models include DeepSeek-V2~\cite{deepseekv2}, DeepSeek-V3~\cite{deepseekv3}, Mixtral 8$\times$7B~\cite{mixtral}, and Qwen2.5-MoE~\cite{qwen25}. 

To better support Mixture-of-Experts (MoE) training on A3 NPUs, we examine its computational structure in depth. Consider the MoE feed-forward network (MoE-FFN) as a representative example. Its forward pass can be decomposed into three high-level stages: (1) dispatching tokens to the ranks that host their selected experts, (2) performing the local expert feed-forward computation, and (3) combining the expert outputs back to the source ranks. As shown in Figure~\ref{fig:dag}(a), these stages are realized by five core operators: \textit{Dispatch}, the first \textit{GroupedGEMM} (\textit{GMM}), \textit{SwiGLU}, the second \textit{GMM}, and \textit{Combine}. \textit{Dispatch} permutes and transfers tokens according to the routing decisions. The first \textit{GMM} computes the gate and up projections, followed by the \textit{SwiGLU} activation, while the second \textit{GMM} performs the down projection.
Finally, \textit{Combine} routes the expert outputs back to the original ranks and accumulates the top-$k$ contributions~\cite{deepseekmoe,deepseekv3}.



The backward pass exposes additional operator-level parallelism. Figure~\ref{fig:dag}(b) contains seven nodes: backward Dispatch,  $\mathrm{GMM}_{\mathrm{act\_grad}}$, $\mathrm{GMM}_{\mathrm{w2\_grad}}$,  $\mathrm{SwiGLU}_{\mathrm{grad}}$, $\mathrm{GMM}_{\mathrm{gate\_grad}}$,  backward Combine, and $\mathrm{GMM}_{\mathrm{w1\_grad}}$.  The backward Dispatch corresponds to the gradient of the forward Combine. Here, $\mathrm{act\_grad}$ computes the intermediate activation gradient, and $\mathrm{gate\_grad}$ computes the gradient along the gate branch, while $\mathrm{w1\_grad}$ and $\mathrm{w2\_grad}$ compute the two expert weight gradients. After backward Dispatch,  $\mathrm{GMM}_{\mathrm{act\_grad}}$ and  $\mathrm{GMM}_{\mathrm{w2\_grad}}$ consume the dispatched  upstream gradient separately. After $\mathrm{SwiGLU}_{\mathrm{grad}}$,  $\mathrm{GMM}_{\mathrm{gate\_grad}}$ and  $\mathrm{GMM}_{\mathrm{w1\_grad}}$ become independent consumers; backward  Combine then returns the resulting input activation gradient~\cite{deepseekv3,mindspore}.


These operators stress different hardware resources.
\textit{GMM} operators mainly use Cube matrix engines, whereas \textit{Dispatch}, \textit{Combine}, \textit{SwiGLU}, activation gradients, and data movement map mostly to AIV resources~\cite{squeezing,accelerating}.

\begin{figure}[!t]
  \centering
  \includegraphics[width=\linewidth]{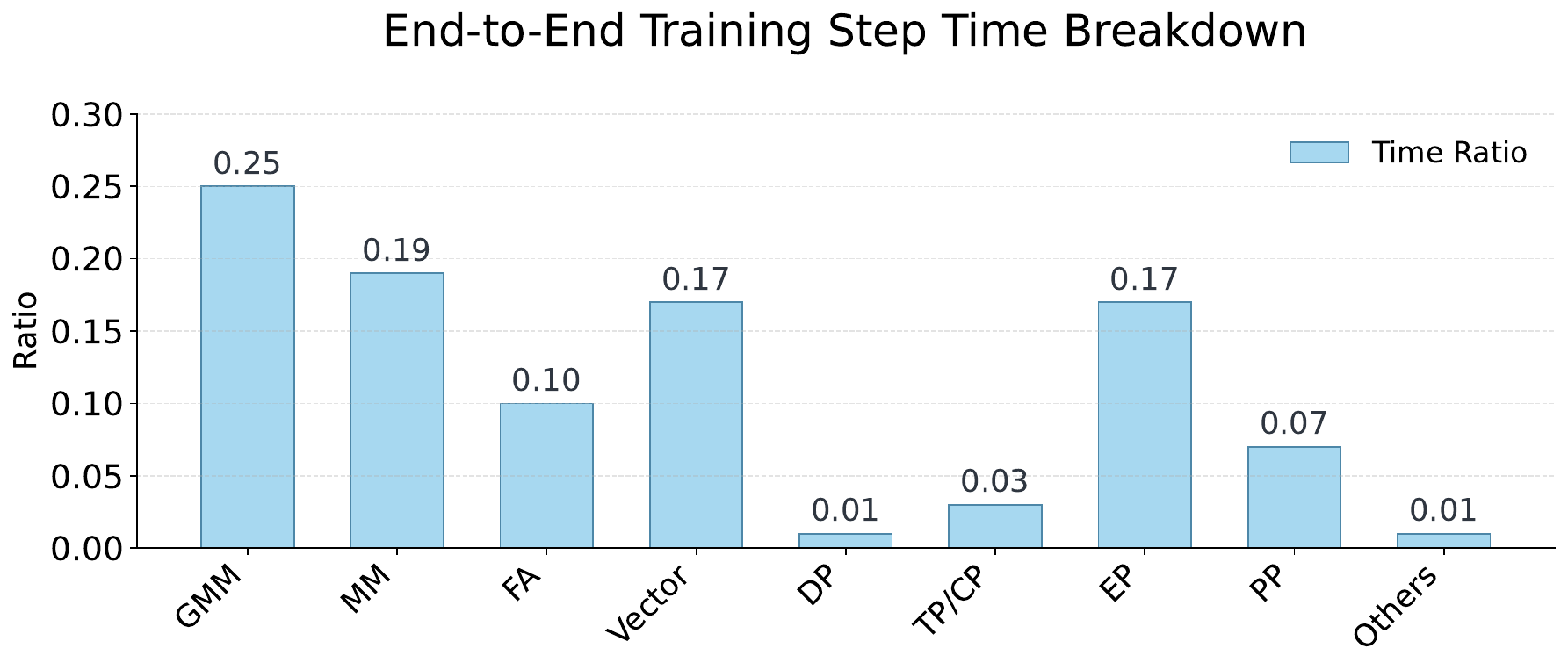}
  \caption{End-to-end training step time breakdown on Ascend A3.}
  \label{fig:breakdown}
\end{figure}

\begin{figure}[t]
  \centering
  \includegraphics[width=\linewidth]{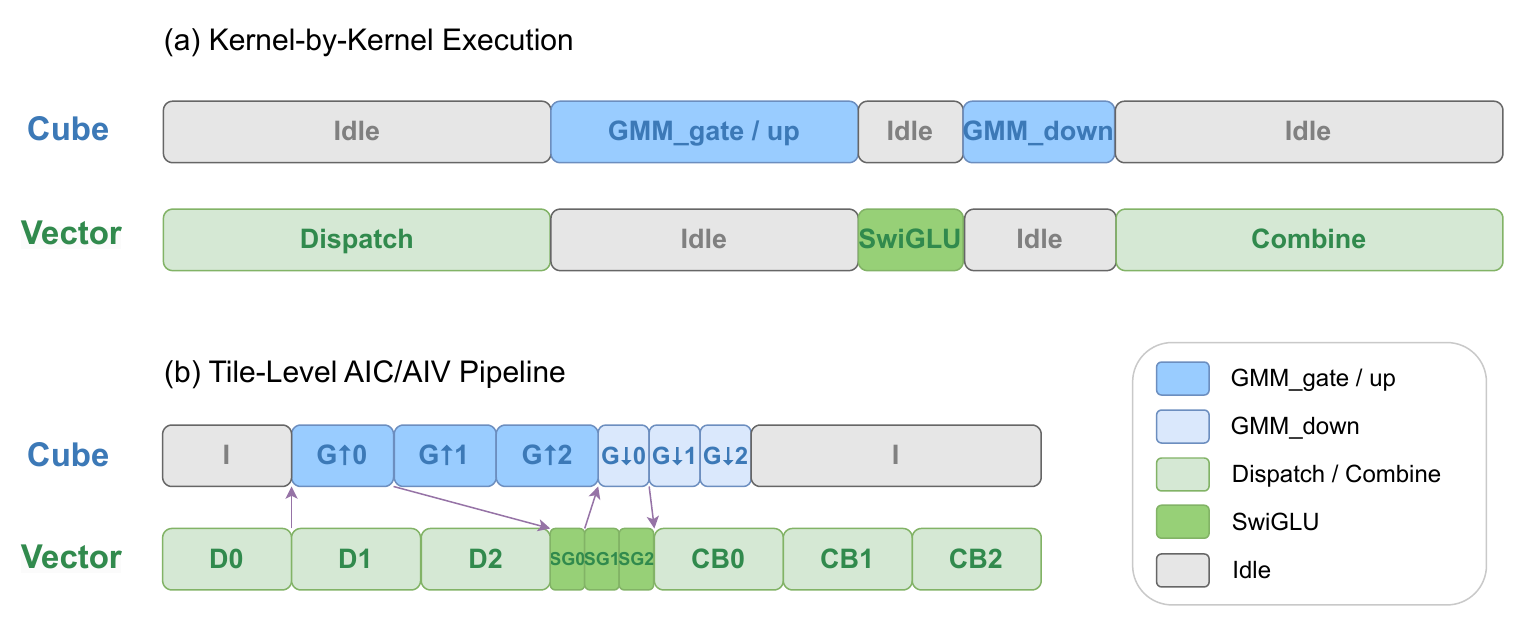}
  \caption{Kernel-by-kernel execution versus tile-level AIC/AIV pipelining.}
  \label{fig:pipeline}
\end{figure}

\subsection{MoE Training Inefficiencies}
\label{sec:waste}

\noindent Efficient training of such MoE models on Ascend A3 NPU clusters demands a thorough understanding of the underlying performance bottlenecks. To this end, we profile a DeepSeek-V3~\cite{deepseekv3} variant with 256 experts, 671B total parameters, and 37B activated parameters, running on a cluster of 256 A3 NPUs (each NPU carries two devices). The training job uses the MindSpore framework~\cite{mindspore} with a parallelism strategy of dp=32, pp=8, tp=2, ep=32. Our profiling reveals several critical performance inefficiencies.

Figure~\ref{fig:breakdown} presents the time breakdown of the major steps in an MoE training job. Matrix computation dominates the overall execution time, with the GroupedGEMM (GMM) operators inside the FFN accounting for the largest share, reaching 25\%. 
To further examine whether the Cube cores are saturated, we use the profiler-reported MAC ratio, defined as the ratio of cycles spent on Cube-type instructions to total cycles.
On average, however, the AIC cores attain only about 67\% MAC ratio. FlashAttention (FA) has been heavily optimized and is no longer a bottleneck~\cite{flashattention}. In contrast, vector operators constitute a non-negligible portion: those that fall on the critical path account for roughly 18\% of the end-to-end step time.

We next analyze the communication overhead. The time costs introduced by data parallelism (DP), tensor parallelism (TP), and context parallelism (CP) have become negligible. However, expert parallelism (EP) imposes a 17\% overhead on the critical path. A closer examination of EP reveals that only about 61\% of its communication latency can be hidden behind Cube computation, leaving roughly 39\% exposed on the critical path. This exposed fraction is expected to grow further as the number of experts increases~\cite{lancet,comet,deepep,megascale}.

Figure~\ref{fig:pipeline}(a) also illustrates the microbatch-level computation pipeline, taking the FFN forward pass as a representative example. We further observe that although Cube and Vector operators use distinct hardware resources and the NPU provides an interface for on-chip heterogeneous scheduling, the pipeline is in fact serial: every kernel is dispatched to the NPU sequentially, causing the AIC and AIV units to idle in alternation rather than operate simultaneously.

The root cause of this problem lies in a mismatch of scheduling patterns. Conventional training stacks, including Megatron-style model-parallel execution~\cite{megatronlm} and the MindSpore baseline used in our profiling~\cite{mindspore}, can overlap coarse computation and communication work, but still launch MoE operators as full-device kernels. They do not expose Ascend's AIC and AIV queues as independent tile-level scheduling targets. As a result, in our profiled Ascend execution path, an AIC-dominant kernel must finish before an AIV-dominant kernel can run, and vice versa, causing the two resource classes to idle in alternation.

Furthermore, in the profiled host-driven EP All-to-AllV (A2AV) path, communication is exposed as an operator-level synchronization point before dependent operators can be launched. This creates an idle interval for dependent AIC and AIV work and motivates fine-grained communication-computation overlap~\cite{comet,lancet,deepep}.

\subsection{Opportunities and Challenges}
\label{sec:opportunity}

\noindent Ascend NPU's physically decoupled heterogeneous on-chip resources and explicit programmability provide a potential optimization path for MoE execution~\cite{squeezing,accelerating}. By decomposing coarse-grained operators into fine-grained tiles and mapping them onto AIC and AIV cores according to computation type, it becomes possible to construct inter-operator pipelines across heterogeneous compute units. As illustrated in Figure~\ref{fig:pipeline}(b), tiles with data dependencies can be organized into pipelined execution flows, while multiple independent streams can further reduce pipeline idle gaps and improve overlap efficiency. However, realizing this vision requires addressing several tightly coupled challenges; inefficiency in any stage may collapse the pipeline into serialized execution or even introduce scheduling overhead exceeding the benefits of overlap.

The first challenge lies in the execution model of AllToAll communication. Conventional collective communication exposes synchronization boundaries, and prior MoE systems show that these boundaries can limit fine-grained communication-computation overlap~\cite{comet,lancet}. Recent GPU communication systems further demonstrate the value of lower-level or device-side communication primitives for EP traffic~\cite{deepep,tritondistributed}. Therefore, \sys\ moves AllToAll progress into AIV-managed asynchronous communication tasks to reduce tile-level communication latency and improve overlap efficiency. Furthermore, because AllToAll tiles must participate in fine-grained runtime scheduling, communication tasks themselves need to be directly managed by AIV cores without host-side orchestration. In this regard, one-sided communication, where AIV independently issues and tracks point-to-point communication operations, naturally matches the design philosophy of the proposed framework.

The second challenge concerns task decomposition under both intra-operator and cross-operator constraints. Within GMM, existing memory reordering and L2 cache reuse optimizations require each tile to cover a complete expert width~\cite{squeezing,accelerating}; overly fine-grained partitioning breaks these optimizations and degrades compute efficiency, whereas overly coarse partitioning increases pipeline bubbles. Across operators, AllToAll tiles and SwiGLU tiles must align with the input and output row partitions of GMM, respectively, to preserve correct tile-level data dependencies. Moreover, AllToAll completion signals must precisely reflect the visibility of destination-side buffers: downstream GMM tiles can only be triggered after the corresponding tile data is fully written and globally visible, preventing premature reads of incomplete data. Consequently, task decomposition is not merely a matter of choosing smaller tile sizes, but rather a tradeoff among intra-operator optimization, inter-operator dependency alignment, synchronization overhead, and pipeline density.

The third challenge is scheduling. After operator decomposition, the system must simultaneously satisfy scheduling correctness, low scheduling overhead, and high scheduling quality. Correctness requires strict enforcement of tile dependencies, such that a GMM tile cannot execute before the corresponding AllToAll tile completes. Low overhead requires dependency tracking and event triggering to remain lightweight; otherwise, scheduling itself becomes a performance bottleneck. High scheduling quality further requires maximizing Cube/Vector overlap and L2 cache reuse under dependency constraints. This introduces a tradeoff between runtime adaptability and critical-path overhead. Online dynamic scheduling can adapt to token distribution changes, but dependency checking, task selection, and queue management all lie on the execution critical path~\cite{pipethreader}. Alternatively, statically generating task orders and synchronization events offline allows runtime execution to simply follow precomputed schedules, reducing runtime decision overhead. The key challenge then becomes generating sufficiently high-quality task sequences while preserving dependency correctness.

Finally, programmability and portability present additional practical challenges. A straightforward solution would be to manually implement fully fused kernels that integrate GMM, SwiGLU, AllToAll, and synchronization logic into a single operator. However, such designs tightly couple scheduling policies, communication protocols, and operator implementations, making them difficult to adapt when model architectures, parallel configurations, data types, or tiling strategies change. Moreover, production-grade GMM, SwiGLU, and communication operators already incorporate substantial hardware-specific optimizations in tiling, memory access, and pipelining~\cite{squeezing,accelerating}; rewriting them from scratch is both costly and likely to sacrifice existing performance optimizations. Therefore, a low-intrusion scheduling framework is needed: existing high-performance operators should be preserved, while dependencies, tiling strategies, and scheduling constraints are declared externally. The framework can then automatically generate task descriptions, queue ordering, and synchronization events, significantly reducing the engineering burden of supporting new MoE variants.

Prior GPU work has shown the value of overlap and pipeline scheduling, but these techniques do not transfer directly to Ascend NPUs. FlashMoE, UniEP, and MegaMoE~\cite{flashmoe,uniep,megamoe} explore persistent or MegaKernel-style MoE execution on NVIDIA GPUs. Triton-distributed~\cite{tritondistributed} exposes distributed primitives in Triton. FLUX, COMET, Lancet, and DeepEP~\cite{flux,comet,lancet,deepep} reduce exposed EP communication, while PipeDream, PipeThreader, Megatron-LM and DHeLlam~\cite{pipedream,pipethreader,megatronlm,dhellam} reduce pipeline bubbles. These systems rely on GPU-specific SMs, compilers, runtimes, or communication kernels, and they do not target Ascend's explicit AIC/AIV queues and cross-queue events. Ascend therefore needs a framework that reuses existing operators while generating AIC/AIV task streams and tile-level synchronization offline.


\begin{figure*}[t]
\centering
\includegraphics[width=0.95\linewidth]{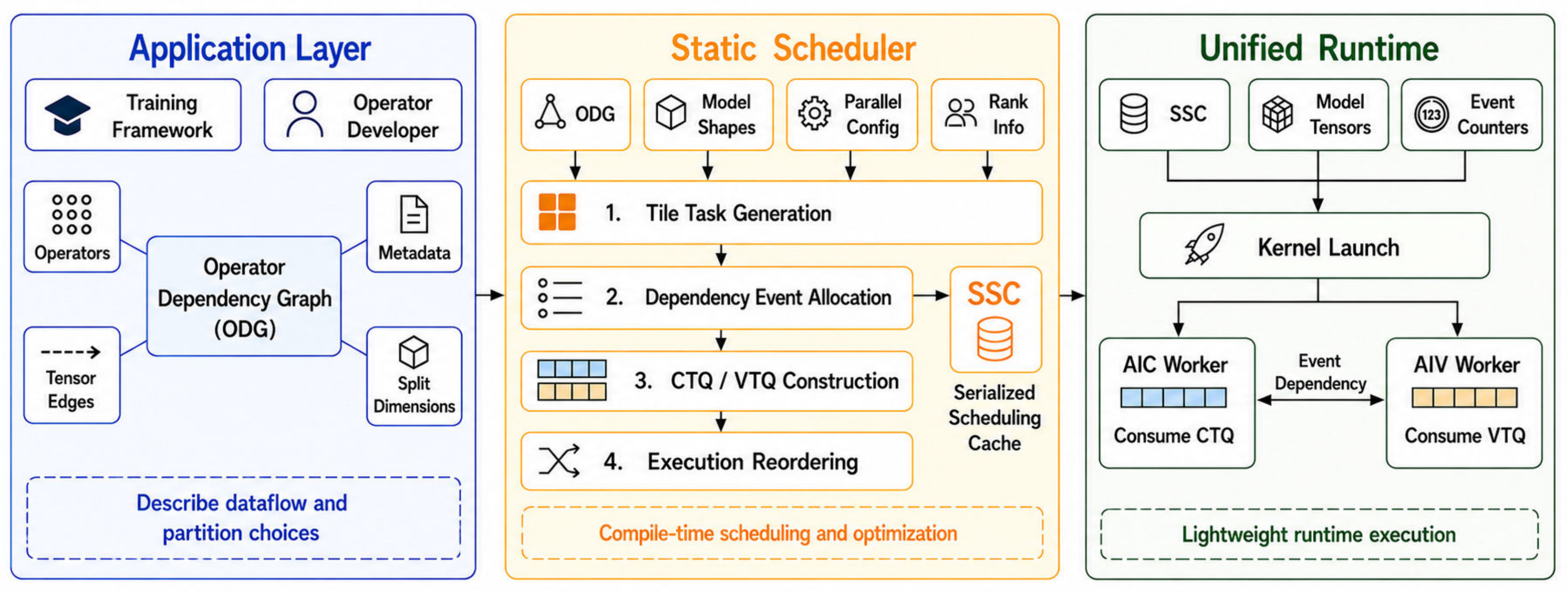}
\vspace{0.8em}
\caption{Overview of \sys.}
\label{fig:overview}
\vspace{1.4em}
\end{figure*}

\section{System Overview}
\label{sect:overview}

To effectively address the challenges discussed above, we present \sys, a compilation and scheduling framework for MoE training on Ascend NPUs that fully exploits the heterogeneous on-chip compute resources. The key idea of \sys\ is to transform the originally operator-serialized MoE-FFN computation graph into a statically scheduled tile-level task pipeline spanning AIC and AIV resources. Instead of executing GMM, SwiGLU, communication, and synchronization as independent kernels separated by global barriers~\cite{comet,lancet,deepep}, \sys\ decomposes them into fine-grained tile tasks and organizes these tasks into concurrent execution streams across heterogeneous hardware queues.

At a high level, \sys\ shifts MoE execution from a \emph{kernel-centric} model to a \emph{taskflow-centric} model. During compilation, the framework analyzes operator dependencies, legal tiling strategies, tensor layouts, and hardware resource types, and determines how each tile task should be mapped onto the underlying execution queues. Compute-intensive matrix tiles are assigned to the Cube Task Queue (CTQ) and executed by AIC workers, while vector computation, data movement, synchronization, and communication progression tasks are assigned to the Vector Task Queue (VTQ) and executed by AIV workers~\cite{squeezing,accelerating}. 
Through this static orchestration, the originally serialized execution across kernel boundaries becomes a unified heterogeneous taskflow capable of concurrent execution across AIC and AIV resources. This design both improves hardware utilization and reduces host-side launch and synchronization overheads.

A central insight behind \sys\ is that communication itself should be elevated to a first-class schedulable task. Rather than treating communication as an externally orchestrated collective operation, \sys\ represents communication progress as tile-level AIV tasks. AIV workers directly drive one-sided communication operations, update synchronization signals, and establish dependencies with AIC tasks through lightweight event counters. As a result, communication progression, data movement, and matrix computation can naturally overlap at tile granularity. Although tasks from different experts, operators, and hardware resources may execute asynchronously and out of order, \sys\ preserves the original MoE computation semantics through explicit dependency events between tasks.

To support this execution model, \sys\ introduces \emph{Static Schedule Configuration} (SSC) as the core abstraction between compilation and runtime. SSC encodes the complete execution plan generated by the static scheduler, including CTQ/VTQ task sequences, task metadata, dependency events, and triggering conditions. Conceptually, SSC serves as a precompiled taskflow that specifies how a particular MoE-FFN instance should execute on heterogeneous hardware resources. Developers only need to declare operator dependency relationships, legal tiling dimensions, and a small amount of operator-specific configuration logic, while the framework automatically generates task queues, synchronization metadata, and execution ordering. This significantly reduces the engineering burden compared with manually implementing fused kernels for every MoE variant~\cite{flashmoe,comet,uniep,megamoe}.

Figure~\ref{fig:overview} illustrates the overall architecture of \sys. The system consists of three major components: the application layer, the static scheduler, and the unified runtime.
The application layer interfaces with upper-level training frameworks and operator developers. It describes an MoE-FFN fragment using an Operator Dependency Graph (ODG), which specifies operators, tensor connections, input/output metadata, and the dimensions along which each operator can be legally partitioned. Importantly, this layer only defines how data flows and how operators may be decomposed, without directly determining execution order.

The static scheduler runs before the training loop begins. Given the ODG, model shapes, parallel configurations, and rank information, it generates tile tasks, allocates dependency events, constructs CTQ/VTQ task streams, and optimizes execution ordering under hardware constraints. 
The resulting execution plan is serialized into SSC for runtime execution.
By moving scheduling decisions offline, \sys\ avoids expensive runtime dependency analysis and task selection on the execution critical path~\cite{pipethreader}.

At runtime, the training framework invokes the unified runtime with cached SSC metadata, model tensors, and event counters. A single kernel launch then executes the corresponding forward or backward taskflow. AIC and AIV workers independently consume CTQ and VTQ tasks while coordinating through lightweight event synchronization. Consequently, runtime logic is reduced to lightweight task fetching, dependency waiting, and event triggering, while complex scheduling decisions remain entirely in the compilation stage.

Through this layered design, \sys\ decouples operator implementation from execution scheduling and enables efficient tile-level heterogeneous pipelining without requiring intrusive rewrites of existing high-performance operators. The detailed mechanisms of operator partitioning, communication taskization, event generation, and queue reordering are presented in Section~\ref{sect:design}.

\section{Design of \sys}
\label{sect:design}

This section presents the design of \sys\ along its execution pipeline. Section~\ref{sec:onesided_design} next first describes how \sys\ transforms EP AllToAll from a host-driven collective communication primitive into schedulable one-sided VTQ tasks. Section~\ref{sec:design_tile} then introduces how dependency-preserving tile tasks are generated from operator graphs, enabling communication, matrix computation, and vector computation to share a unified task abstraction. Section~\ref{sec:design_schedule} further explains how the static scheduler compiles these tasks into CTQ/VTQ taskflows augmented with queue ordering and event-driven synchronization semantics. Section~\ref{sec:design_runtime} describes how the unified runtime drives concurrent AIC/AIV execution within a single kernel launch. Finally, Section~\ref{sec:design_order} presents how \sys\ optimizes legal task ordering under dependency constraints to improve cross-rank communication balance and L2 cache locality.

\subsection{AIV-Driven One-Sided Communication}
\label{sec:onesided_design}

\sys\ builds its communication mechanism upon device-side remote memory writes and device-side signal updates. Ascend NPUs provide SHMEM-like remote memory access capabilities~\cite{squeezing,accelerating}, where an AIV worker on one rank can directly access communication buffers exposed by a remote rank through device-side address translation, and can further update synchronization signals on the destination side. \sys\ leverages this capability to move communication from the host-side collective communication runtime into the device-side AIV execution path. As a result, communication is no longer treated as a globally synchronized operator stage initiated by the host, but instead becomes an actively progressing component within the unified taskflow executed by AIV workers.

Based on this capability, \sys\ encapsulates device-side remote write with completion notification into a communication primitive called \texttt{put\_mem\_signal}. The interface takes as input a source buffer, destination rank, destination buffer, transfer range, and a destination-side event counter. 
Its semantics are straightforward: the sender-side AIV worker writes a specified data block into the remote buffer of the destination rank, and updates the destination-side event counter once the transferred data becomes globally visible to the receiver. 
Conventional communication primitives return tensors. In contrast, this primitive exposes two observable side effects on the destination rank: the target buffer becomes readable, and the corresponding event counter is updated. Therefore, it simultaneously serves as both a data transfer operation and a completion notification mechanism.

Using this primitive, \sys\ decomposes the AllToAll communication into a collection of fine-grained  tasks of \texttt{put\_mem\_signal}. This taskization follows the same high-level goal as prior systems that expose communication at finer granularity or through device-side communication primitives~\cite{comet,deepep,tritondistributed}, but the completion semantics here are the Ascend AIV-side remote-write and event-counter semantics described above. Each AllToAll tile corresponds to one remote write operation that transfers a contiguous data block from a source rank to a destination rank. The communication task descriptor records metadata including source tensor, destination tensor, destination rank, communication range, and synchronization events. During execution, the AIV worker computes the source offset, destination offset, and transfer size for the current tile, and invokes \texttt{put\_mem\_signal} to complete both data transfer and signal update. Since each communication tile independently updates its associated completion counter after data becomes visible on the destination side, downstream computation can determine data readiness based on fine-grained tile completion rather than waiting for the entire AllToAll phase to finish.

Dispatch AllToAll and Combine AllToAll correspond to two opposite directions of this communication mechanism. During Dispatch, source ranks send routed tokens to the destination ranks that host the selected experts. \sys\ decomposes each expert/rank communication region into communication tasks. Once all required input tiles for a local expert have arrived on the destination rank and their completion signals indicate readability, the corresponding GMM task can begin execution immediately without waiting for Dispatch completion of unrelated experts or ranks.

During Combine, the communication direction is reversed. After expert computation completes, destination ranks send expert outputs back to the original token source ranks, where top-$k$ expert contributions are accumulated according to routing scores. Combine communication tasks depend on the completion events of their corresponding local expert output tiles. Once a GMM task produces a complete and readable output tile, the associated Combine task can invoke \texttt{put\_mem\_signal} to return the result to the source rank and update the source-side completion counter.

Through this transformation, AllToAll communication evolves from a host-side opaque collective communication stage into a collection of VTQ tasks with explicit inputs, outputs, and completion events. Dispatch and Combine communication tasks follow the same execution model as ordinary AIV tasks such as SwiGLU, allowing the scheduler to jointly orchestrate communication, vector computation, and synchronization within a unified taskflow. More importantly, communication tasks can establish explicit dependencies with CTQ-resident GMM tasks through event counters, enabling communication, matrix computation, and vector computation to coexist within a single dependency graph and be statically scheduled together at tile granularity.

\begin{algorithm}[t]
\caption{Split Propagation}
\label{alg:split-propagation}
\footnotesize
\begin{algorithmic}[1]
\Statex \textbf{Input:} ODG $G$, shape and parallel configuration $C$
\Statex \textbf{Output:} Task count $op.\textit{task\_num}$ for each operator, and split labels on output tensors
\Statex \textit{// Split labels are stored on tensors shared by producer outputs and consumer inputs.}
\ForAll{tensor $x$ in $G$}
    \State $x.\textit{split\_dim} \gets -1$
    \State $x.\textit{split\_num} \gets 1$
\EndFor
\ForAll{operator $op$ in $\textsc{TopologicalSort}(G)$}
    \State $s \gets op.\textit{SplitSpec}$
    \If{$s.\textit{split\_inputs} = \texttt{None}$}
        \State $n \gets s.\textit{task\_num\_fn}(C)$
    \ElsIf{$\forall (i,d) \in s.\textit{split\_inputs},\
        op.\textit{inputs}[i].\textit{split\_dim} = d$}
        \State $n \gets s.\textit{task\_num\_fn}(C)$
    \Else
        \State $n \gets 1$ \Comment{fallback to one unsplit task}
    \EndIf
    \State $op.\textit{task\_num} \gets n$
    \ForAll{output $y_j \in op.\textit{outputs}$}
        \State $d \gets s.\textit{split\_output\_dims}[j]$
        \If{$n > 1$ and $d \ge 0$}
            \State $y_j.\textit{split\_dim} \gets d$
            \State $y_j.\textit{split\_num} \gets n$
            \Comment{visible to downstream inputs}
        \Else
            \State $y_j.\textit{split\_dim} \gets -1$
            \State $y_j.\textit{split\_num} \gets n$
        \EndIf
    \EndFor
\EndFor
\end{algorithmic}
\end{algorithm}

\subsection{Tile Task Generation from Operator Graphs}
\label{sec:design_tile}

In \sys, tile tasks serve as the fundamental execution unit. Each tile task must explicitly specify its target execution queue, input and output tensor ranges, and dependency endpoints with upstream and downstream tasks. However, tile boundaries cannot be chosen arbitrarily: overly fine-grained or misaligned partitioning may break the carefully optimized tiling, memory access, and buffering strategies in production-grade operators~\cite{squeezing,accelerating,deepgemm}, and may even expose incomplete data to downstream consumers. Therefore, \sys\ determines tile task boundaries during static compilation rather than runtime execution.

Instead of generating tasks independently for each operator, \sys\ constructs tile tasks from an Operator Dependency Graph (ODG), enabling cross-operator aligned task decomposition. ODG is the scheduling intermediate representation introduced by \sys\ to describe the operator-level dataflow within a schedulable computation fragment. Nodes in the graph correspond to \texttt{OperatorNode}s, while edges represent tensor dependencies between operators. Each \texttt{OperatorNode} records operator type, input/output tensors, and an associated \texttt{SplitSpec} that defines the legal tiling strategy for the operator.

Specifically, \texttt{SplitSpec} captures three aspects of task decomposition: which input tensor partition should be inherited by the current operator, along which dimension the output partition should continue propagating downstream, and how many tile tasks should be generated under a given tensor shape and parallel configuration. These decomposition rules must simultaneously satisfy both operator semantics and hardware execution constraints.

For GMM operators, \sys\ only exposes task-level parallelism along independent token or expert-block dimensions, while preserving reduction dimensions such as $K$ intact. This design avoids breaking the internal accumulation structure and expert-local output layout, thereby preserving existing GMM optimizations including expert grouping, memory reordering, and L2 cache locality~\cite{squeezing,accelerating,deepgemm}. For AIV-side operators, including Dispatch, SwiGLU, and Combine, tile boundaries must align with the input or output row partitions of GMM operators. Consequently, every downstream task consumes a well-defined tensor region produced by a deterministic set of upstream producers.

Declaring legal decomposition rules for individual operators alone is insufficient, because downstream operators can only generate compatible tile tasks when their inputs already carry the expected partitioning structure. To address this issue, \sys\ performs split propagation over the ODG to infer task counts and output partition labels across operators. Split propagation traverses the graph in topological order and applies the \texttt{SplitSpec} of each node sequentially.

For operators that do not inherit upstream partitioning, such as Dispatch, \texttt{SplitSpec} directly generates the initial decomposition. For downstream operators, tile tasks are generated only when their inputs already carry the required partition labels. Otherwise, the operator falls back to a non-partitioned execution mode to preserve semantic correctness at the cost of reduced parallelism. Algorithm~\ref{alg:split-propagation} illustrates this process. Conceptually, split propagation is implemented through tensor-level partition annotations: upstream operators write partition labels into output tensors, while downstream operators retrieve these labels when validating their \texttt{split\_inputs} requirements.

As an example, consider a forward MoE-FFN pipeline. Dispatch specifies \texttt{split\_inputs=None}, meaning that it does not inherit partitioning from upstream operators. Instead, it serves as the partitioning origin and generates Dispatch communication tasks while annotating its output tensors with row-wise partition labels. This row partitioning is then propagated through the main dataflow path across GMM1, SwiGLU, and GMM2, allowing subsequent compute tasks to be generated over compatible row partitions. Combine additionally consumes communication metadata such as offset and size tensors; therefore, it ignores these metadata inputs during split checking and instead inherits row-wise partitioning from the source-data input produced by GMM2. Based on this inherited partitioning, Combine generates communication tasks aligned with GMM2 outputs. Through this propagation mechanism, communication, matrix computation, and vector computation all obtain compatible task boundaries before execution scheduling begins.

\begin{table}[t]
  \centering
  \caption{Key fields in a Task Descriptor (TD).}
  \label{tab:td}
  \scriptsize
  \begin{tabularx}{\linewidth}{@{}P{23mm}P{14mm}L@{}}
    \toprule
    \textbf{Field} & \textbf{Type} & \textbf{Semantics} \\
    \midrule
    \texttt{task\_type}
      & \texttt{uint32}
      & Task handler type, e.g., GMM, SwiGLU, SwiGLUGrad, or \texttt{put\_mem\_signal} \\
    \texttt{queue\_type}
      & \texttt{uint32}
      & Target execution queue: CTQ for AIC tasks or VTQ for AIV tasks \\
    \texttt{dependent\_event}
      & \texttt{uint32}
      & Event counter that must reach its threshold before this task starts \\
    \texttt{trigger\_event}
      & \texttt{uint32}
      & Event counter updated after this task completes \\
    \texttt{inputs}
      & \texttt{TDesc[]}
      & Input tensor slots, shapes, flags, and tile offsets \\
    \texttt{outputs}
      & \texttt{TDesc[]}
      & Output tensor slots, shapes, flags, and tile offsets \\
    \texttt{task\_index}
      & \texttt{uint32}
      & Tile index within the operator \\
    \texttt{task\_split\_num}
      & \texttt{uint32}
      & Number of tile tasks generated for this operator \\
    \texttt{task\_split\_value}
      & \texttt{uint32}
      & Per-task split granularity used to derive tile ranges \\
    \shortstack[l]{\texttt{tiling\_data\_}\\\texttt{position}}
      & \texttt{uint32}
      & Argument-list position of the operator tiling data \\
    \bottomrule
  \end{tabularx}
\end{table}

To materialize the runtime metadata required for tile execution, \sys\ provides an operator-specific \texttt{FillConfig} for each operator type. \texttt{FillConfig} defines how legal tile tasks are transformed into runtime-consumable Task Descriptors (TDs). A TD is the basic runtime-consumed task unit and records the task type, hardware resource type, input/output tensor ranges, task identifiers, split strides, tiling pointers, and operator-specific metadata. Table~\ref{tab:td} summarizes the key TD fields.

After \texttt{FillConfig} converts tile tasks into TDs, operator-specific details such as offset computation, tiling selection, and private metadata layouts become encapsulated within a unified task representation. Consequently, the static scheduler only needs to reason about TD-level resource types and dependency endpoints when constructing CTQ and VTQ schedules, without reanalyzing operator internals.

Through ODG-driven task generation, \sys\ ultimately produces a static task collection with dependency-preserving semantics, aligned tile boundaries, and explicit resource assignments. Dispatch and Combine communication tasks, GMM tasks, and SwiGLU tasks are all represented within the same task abstraction. This design avoids the coarse-grained synchronization boundaries introduced by operator-level execution while simultaneously preventing excessively fine-grained decomposition from disrupting existing operator-level optimizations. The next section describes how \sys\ further organizes these TDs into CTQ and VTQ schedules and generates low-overhead static execution plans through event-based synchronization.

\subsection{Static Scheduling and Event-Driven Synchronization}
\label{sec:design_schedule}

The TD collection generated in the previous stage specifies how each tile task should execute, but does not yet determine when tasks should execute. The scheduler must further determine two aspects of execution: the placement of each task within CTQ or VTQ, and the synchronization relationships among tasks. The former defines the execution order candidates within the same resource type, while the latter ensures that AIC and AIV workers do not violate true data dependencies during concurrent execution.

\begin{table}[t]
  \centering
  \caption{Static and dynamic scheduling tradeoffs for \sys\ taskflows.}
  \label{tab:static_dynamic}
  \scriptsize
  \begin{tabularx}{\linewidth}{@{}P{24mm}P{31mm}L@{}}
    \toprule
    \textbf{Aspect} & \textbf{Static scheduling} & \textbf{Dynamic scheduling} \\
    \midrule
    Runtime work
      & Consumes precomputed CTQ/VTQ and event metadata
      & Builds or updates ready queues during execution \\
    Critical-path cost
      & Performs task fetch, event wait, and event trigger
      & Performs readiness checks and task selection online \\
    Dependency handling
      & Encodes dependencies before launch using event counters
      & Resolves dependencies from runtime task state \\
    Ordering policy
      & Applies communication- and locality-aware ordering offline
      & Adapts ordering to runtime state with online overhead \\
    Shape variability
      & Reuses SSC for stable shapes or shape buckets
      & Handles new shapes after runtime metadata is available \\
    Implementation fit
      & Matches Ascend CTQ/VTQ execution with lightweight workers
      & Requires a device-side scheduler and mutable task state \\
    \bottomrule
  \end{tabularx}
\end{table} 

\sys\ adopts static scheduling to make these decisions offline. During stable training phases, MoE models typically exhibit fixed operator structures, parallel configurations, and dominant tensor shapes. As a result, task decomposition, resource mapping, and dependency topology can all be determined before execution begins. In contrast, online dynamic scheduling requires runtime task selection, dependency checking, and queue management on the device execution path, introducing overhead on the critical path and consuming hardware resources that could otherwise be used for computation or communication~\cite{pipedream,pipethreader,comet}. By moving scheduling decisions into the compilation stage, \sys\ reduces runtime responsibilities to lightweight task fetching, dependency waiting, and event triggering. Table~\ref{tab:static_dynamic} summarizes this tradeoff in \sys's target setting.

The scheduler first constructs two execution queues according to the resource type recorded in each TD. Matrix-intensive tasks such as GMM are placed into CTQ and executed by AIC workers, while Dispatch, Combine, SwiGLU, and other vector or communication-related tasks are placed into VTQ and executed by AIV workers~\cite{squeezing,accelerating}. Importantly, queue order only defines execution ordering among tasks of the same resource type. Cross-resource dependencies are not implicitly enforced through queue position, but instead explicitly represented through event synchronization. Consequently, CTQ and VTQ can progress independently while still preserving semantic correctness through dependency events.

\sys\ uses event counters as the synchronization mechanism between heterogeneous queues. Each task is associated with two events: a \emph{dependent event}, representing the prerequisite completion condition required before execution, and a \emph{trigger event}, representing the completion signal that will be emitted after execution finishes. Each event is additionally associated with a statically generated threshold indicating how many times the event must be triggered before dependent downstream tasks become executable.

Before executing a task, the worker continuously polls the counter associated with the dependent event until the counter value reaches the required threshold. After task completion, the worker updates the counter associated with the trigger event, thereby releasing downstream tasks waiting on that condition. Event counters reside in device-visible memory and are accessible to both AIC and AIV workers~\cite{squeezing,accelerating}. Since all event relationships and thresholds are statically generated during compilation, runtime execution does not require dependency graph construction or complex scheduling logic.

This threshold-based synchronization mechanism naturally expresses many-to-many tile-level dependencies. Multiple upstream tasks may contribute to the same event counter, and downstream execution begins only after the accumulated counter value reaches the corresponding threshold. For example, a GMM task may wait until several Dispatch communication tiles have all arrived before computation begins. Conversely, multiple downstream tasks may wait on the same event, thereby sharing a common completion condition. For instance, completion of a GMM output tile may simultaneously release subsequent vector computation tasks and communication tasks. Through this design, \sys\ synchronizes based on true tile-level data readiness rather than coarse-grained operator-level barriers.

Scheduling correctness follows from two properties. First, all tasks are generated from the ODG and \texttt{SplitSpec}, ensuring that each task has explicitly defined input regions and deterministic upstream producers. Second, events are triggered only after upstream outputs become fully readable, while downstream tasks remain blocked until event counters reach the required thresholds. Consequently, although \sys\ changes task granularity and execution ordering, it preserves the original data dependencies and mathematical semantics of the MoE-FFN computation graph.

Moreover, since the ODG is a directed acyclic graph, split propagation only proceeds along topological order, and statically generated event relationships never introduce cyclic dependencies. Workers simply consume tasks sequentially from CTQ and VTQ while respecting event waits, guaranteeing continuous forward progress of the taskflow.

After static scheduling, \sys\ produces a CTQ/VTQ taskflow augmented with explicit execution ordering and synchronization semantics, and serializes this execution plan into SSC for runtime consumption. The next section describes how the unified runtime interprets this taskflow within a single kernel execution window and drives concurrent execution across AIC and AIV workers.

\subsection{Unified AIC/AIV Task Execution}
\label{sec:design_runtime}

Unified AIC/AIV task execution refers to the execution model in which \sys\ concurrently advances both AIC and AIV tasks within the same device-side execution window. Unlike traditional kernel-by-kernel execution, where Dispatch, GMM, SwiGLU, and Combine are launched sequentially as independent kernels, \sys\ executes an entire MoE-FFN forward or backward taskflow within a single unified kernel launch~\cite{mindspore,megatronlm,flashmoe,uniep,megamoe}. Under this model, AIC workers consume matrix-computation tasks from CTQ, while AIV workers consume vector-computation, data-movement, and communication tasks from VTQ. The two worker groups synchronize through event counters and progress concurrently at tile granularity.

To realize this execution model, the training framework passes SSC, model tensors, event counters, and necessary runtime metadata into the unified runtime at each invocation. The runtime launches both AIC workers and AIV workers within the same kernel context. Although both worker types follow the same queue-consumption protocol, they fetch tasks from different queues. Each worker retrieves the next TD index from CTQ or VTQ according to its resource type, loads the corresponding TD, waits for dependency satisfaction, and dispatches execution to the appropriate task handler according to the \texttt{task\_type} field recorded in the TD, such as GMM handlers, SwiGLU handlers, or communication handlers. Importantly, the runtime neither modifies queue order nor reallocates synchronization events; it simply executes the precompiled CTQ/VTQ taskflow encoded in SSC.

Task handlers further bridge the scheduling framework and existing optimized operator implementations. TDs specify the input/output tensor regions and operator-specific metadata associated with each task, while handlers invoke the corresponding GMM, SwiGLU, Combine, or communication logic based on this metadata. \sys\ therefore does not require rewriting the computational core of production-grade operators. Instead, existing operators are encapsulated into taskflow-compatible execution handlers. This design cleanly decouples scheduling policy from operator implementation: the scheduler determines task granularity and execution order, while handlers preserve operator semantics during task execution.

This execution model directly enables overlap between AIC and AIV activities. Consider a forward MoE-FFN pipeline. Under conventional execution, Dispatch must complete globally before the first GMM kernel launches, and the first GMM stage must fully finish before SwiGLU begins, with subsequent stages following the same serialized pattern~\cite{comet,lancet,flux,deepep}. In contrast, \sys\ allows a GMM task for a particular expert to begin execution immediately once its required token block arrives, while AIV workers continue advancing Dispatch communication for other experts or executing SwiGLU and Combine tasks for experts that have already completed computation. As a result, Cube computation for one expert, Vector computation for neighboring experts, and communication progression can all overlap within the same execution window.

Beyond increasing concurrency, unified execution also exposes on-chip data reuse opportunities as a schedulable optimization target. Since AIC and AIV share the same L2 cache~\cite{squeezing,accelerating}, outputs produced by GMM tasks may still reside in L2 when consumed by downstream AIV tasks. If the scheduler places producers and consumers close in execution time, downstream AIV tasks can directly reuse intermediate results from L2 rather than reloading them from HBM. Importantly, this benefit does not require modifying operator internals; instead, it emerges naturally from exposing task granularity and execution ordering to the scheduler.

Therefore, Unified AIC/AIV task execution serves as the runtime foundation of \sys. It enables AIC and AIV workers to cooperatively execute precompiled heterogeneous taskflows within a single kernel lifetime, while creating opportunities for concurrent execution and on-chip cache reuse. However, even under dependency constraints, substantial freedom still exists in determining legal task orderings. The next section describes how \sys\ exploits this flexibility to further optimize communication balance and cache locality.

\subsection{Execution-Order Optimizations}
\label{sec:design_order}

Static scheduling must first guarantee correctness, but a correct execution order is not necessarily efficient. Within the same dependency graph, many tasks do not have direct dependencies and therefore admit multiple semantically equivalent legal execution orders. Different orderings can significantly affect cross-rank communication balance, inter-rank waiting time, and opportunities for L2 cache reuse~\cite{comet,pipethreader,squeezing}. \sys\ formulates execution-order optimization as the problem of selecting efficient task schedules within the space of legal topological orderings.

\sys\ reorders only tasks that are mutually independent. The scheduler does not modify ODG edges, task input/output regions, or event-trigger semantics. Instead, it only changes the relative order of independent tasks within the same queue, or selects a more efficient ordering among multiple legal topological schedules. This constraint guarantees that execution-order optimization preserves program semantics while still exposing meaningful optimization opportunities.

\begin{figure}[t]
  \centering
  \includegraphics[width=\linewidth]{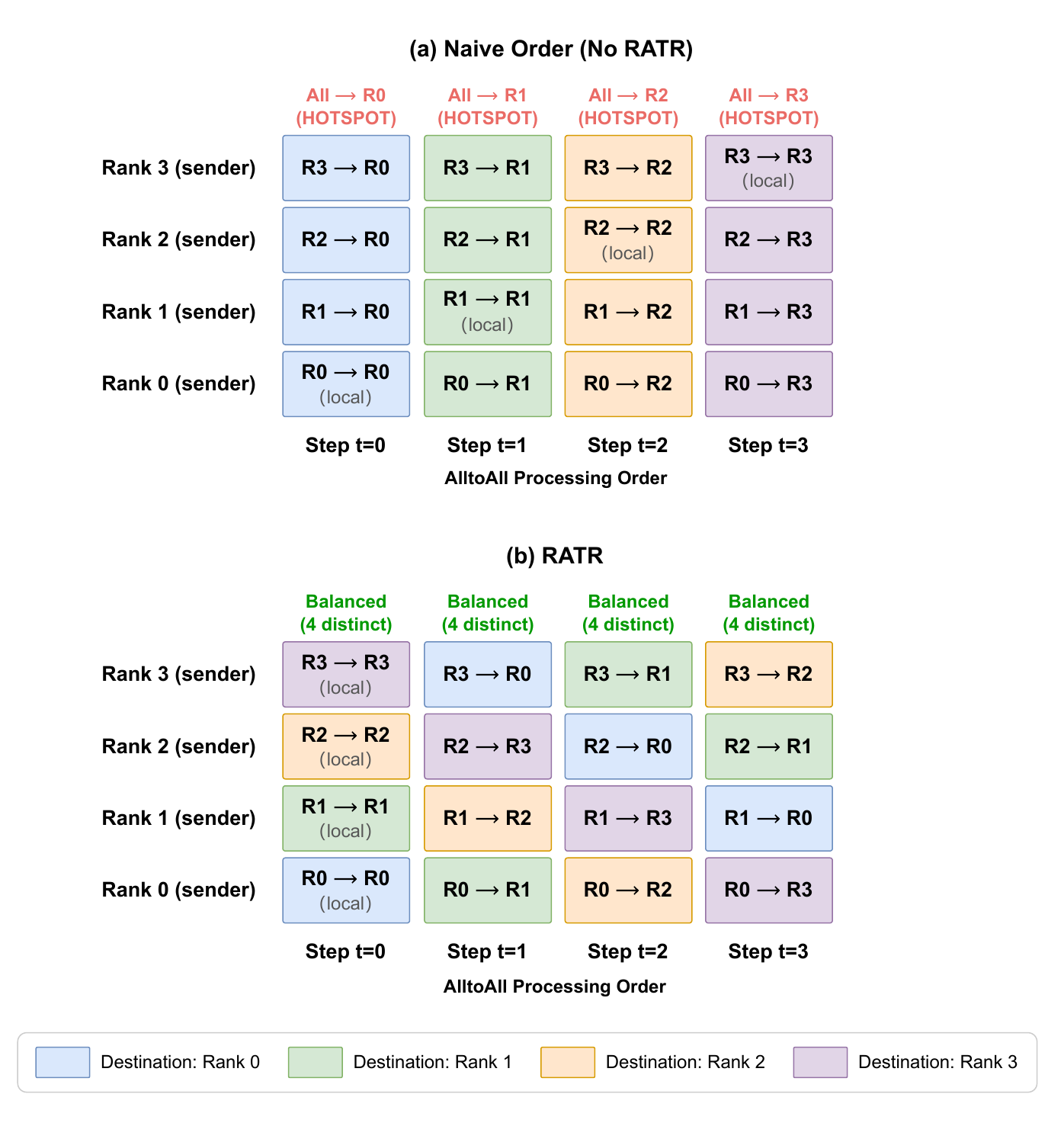}
  \caption{Rank-Aware Task Reordering (RATR).  The naive order creates
  destination-rank hotspots, while RATR rotates each rank's task order to form
  a balanced communication pattern.}
  \label{fig:ratr}
\end{figure}

The first optimization is \emph{rank-aware task reordering} (RATR). For EP AllToAll, a naive order makes all ranks issue communication tasks in the same destination-rank sequence. As a result, communication traffic may become concentrated on the same destination ranks or communication links during certain time windows, leading to communication imbalance and increasing completion-time skew across ranks~\cite{comet,deepep,megascale}. To alleviate this issue, \sys\ rotates the communication-task ordering according to source rank ID, as illustrated in Figure~\ref{fig:ratr}. Different source ranks therefore begin communication from different destination ranks and traverse the remaining destinations in a ring-like order. This transformation does not change the set of transmitted data or dependency relationships, but only changes the relative ordering of independent communication tasks. Consequently, communication traffic becomes temporally distributed across different destinations, improving load balance within the EP group and reducing downstream waiting caused by a small number of lagging ranks.

The second optimization is \emph{cache-guided GMM interleaving}. In backward MoE-FFN execution, multiple GMM branches may consume shared input activations while remaining topologically independent. For example, after backward Dispatch, both the activation-gradient GMM and the down-projection weight-gradient GMM consume the dispatched expert activations without depending on each other. If the scheduler executes one GMM branch in its entirety before launching the other, shared expert activations may already be evicted from L2 cache before reuse occurs~\cite{squeezing,accelerating,deepgemm}. \sys\ therefore interleaves these GMM tasks according to expert locality, placing GMM tasks associated with the same expert as closely together in time as possible. This optimization does not modify dependencies between the two GMM branches, nor does it alter the input/output semantics of individual GMM tasks. Instead, it only changes the relative ordering of independent CTQ-resident GMM tasks, thereby shortening the reuse interval of shared activation data and increasing opportunities for L2 cache reuse.

Overall, these optimizations stay within legal topological orders. RATR balances traffic across ranks, and GMM interleaving improves L2 cache reuse. Neither optimization changes task semantics or event dependencies, allowing \sys\ to improve static taskflow efficiency while preserving correctness.

\section{Evaluation}
\label{sect:eval}

\subsection{Implementation Details}
\label{sec:eval_implementation}

We implement \sys\ in the MindSpore~\cite{mindspore} and MindFormers~\cite{mindformers} MoE-FFN path on Ascend A3~\cite{squeezing,accelerating}. A host-side compiler constructs the ODG, applies legal split rules, materializes TDs, assigns event counters, orders task queues, and serializes the per-rank result into a Static Schedule Configuration (SSC). Compilation is outside the timed module interval. For a fixed shape bucket, EP size, and rank, the SSC is reused across training steps; each step supplies only tensor pointers, routing-derived offsets, and fresh event-counter state.

\para{Runtime integration.} During runtime, MindSpore~\cite{mindspore} launches one unified AscendC operator~\cite{squeezing,accelerating} with model tensors, the SSC, communication buffers, and event counters. AIC workers consume CTQ entries, AIV workers consume VTQ entries, and both worker groups follow a simple protocol: fetch a precompiled TD, wait on its dependency counter, execute the selected handler, and trigger successor counters. The runtime does not build dependency graphs, allocate TDs, or reorder queues on the critical path.

\para{Operator reuse and validation.} Task handlers wrap the existing optimized compute bodies for GMM, SwiGLU, and related MoE-FFN operators, while Dispatch and Combine use AIV-side \texttt{put\_mem\_signal} operations over preallocated device-visible buffers. Unless otherwise stated, \sys\ denotes the full optimized configuration, including one-sided communication, Rank-Aware Task Reordering (RATR), and cache-guided GMM interleaving for backward execution. We validate forward and backward bf16 outputs against the standard MindSpore MoE-FFN path~\cite{mindspore,mindformers,deepseekv3} before timing.

\subsection{Experimental Setup}
\label{sec:eval_setup}
We evaluate \sys\ at two levels. The module-level experiment measures the MoE-FFN boundary transformed by the proposed scheduler: the interval starts at Dispatch and ends after Combine, and includes the unified AscendC runtime, \texttt{put\_mem\_signal} communication, event-counter synchronization, RATR, and backward GMM interleaving. We also report end-to-end training-step latency to show how the optimized MoE-FFN path affects a complete step; this interval includes unchanged model computation and framework overhead in addition to the transformed MoE-FFN execution.

Experiments run on 64 Ascend A3 devices using the MindSpore~\cite{mindspore} and MindFormers~\cite{mindformers} training stack. Each Ascend A3 device provides 25 AIC units, 50 AIV units, and a 192 MB L2 cache~\cite{squeezing,accelerating}. The workload follows a DeepSeek-V3-style MoE-FFN module~\cite{deepseekv3} with sequence length 4096, microbatch size 2, hidden size 7168, intermediate size 2048, top-$k=8$, and bf16 arithmetic. We evaluate EP group sizes 4, 8, and 16, corresponding to 32, 64, and 128 total experts with 8 local experts per rank. The corresponding training configuration labels are DP32/TP2/EP4, DP32/TP2/EP8, and DP32/TP2/EP16, all with global batch size 512. Module latency is measured at the Dispatch-to-Combine boundary, while step latency is measured around one complete training iteration. The baseline is the standard operator-by-operator execution path with full-device operators, full-core exclusive execution, and collective AllToAll communication. For end-to-end step latency, the baseline also retains MindSpore's DVM-level automatic fusion and graph-level execution planning, which can partially mask the module-level gains.

We report two complementary measurements after training reaches steady-state execution. The primary controlled setting uses balanced routing and reports module latency at the Dispatch-to-Combine boundary. The step-level setting uses sampled natural routing and reports end-to-end training-step latency.

\subsection{Main Results: MoE-FFN Module and Step Latency}
\label{sec:eval_main_results}

We first evaluate Dispatch-to-Combine MoE-FFN latency under a balanced-routing setting. Each expert receives the same number of tokens, so the comparison isolates the effect of tile-level scheduling, one-sided communication, and execution-order optimizations from router-induced skew. We then report end-to-end training-step latency under sampled natural routing, which tests whether the module-level gain remains visible after including the unchanged parts of the training stack.

\begin{table}[t]
  \centering
  \caption{Dispatch-to-Combine MoE-FFN latency with balanced routing.}
  \label{tab:balanced_latency}
  \scriptsize
  \begin{tabularx}{\linewidth}{@{}LP{13mm}rrr@{}}
    \toprule
    \textbf{EP setting} & \textbf{Path}
      & \textbf{Baseline (ms)} & \textbf{\sys\ (ms)}
      & \textbf{Speedup} \\
    \midrule
    EP4               & Forward  & 16.3 & 10.2 & 1.60$\times$ \\
                      & Backward & 27.9 & 19.4 & 1.44$\times$ \\
                      & Total    & 44.2 & 29.6 & 1.49$\times$ \\
    \midrule
    EP8               & Forward  & 17.3 & 10.3 & 1.68$\times$ \\
                      & Backward & 29.8 & 19.6 & 1.52$\times$ \\
                      & Total    & 47.1 & 29.9 & 1.58$\times$ \\
    \midrule
    EP16              & Forward  & 18.4 & 11.2 & 1.64$\times$ \\
                       & Backward & 30.5 & 19.9 & 1.53$\times$ \\
                       & Total    & 48.9 & 31.1 & 1.57$\times$ \\
    \bottomrule
  \end{tabularx}
\end{table}

\begin{figure}[t]
  \centering
  \includegraphics[width=0.8\linewidth]{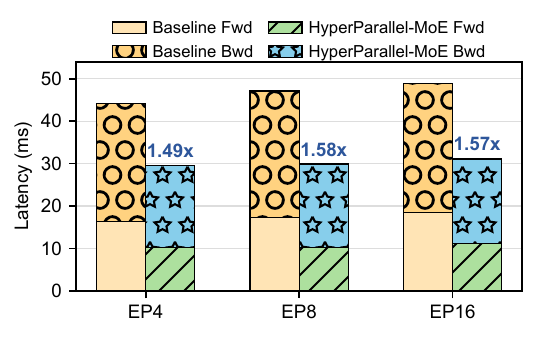}
  \caption{Forward/backward Dispatch-to-Combine latency breakdown under balanced routing. Bar annotations report total speedup over the standard operator-by-operator baseline.}
  \label{fig:balanced_latency_breakdown}
\end{figure}

\para{Balanced routing.} Table~\ref{tab:balanced_latency} and Figure~\ref{fig:balanced_latency_breakdown} report forward, backward, and total Dispatch-to-Combine latency under forced load-balanced routing. Across EP4, EP8, and EP16, \sys\ consistently reduces total module latency from 44.2--48.9 ms to 29.6--31.1 ms, corresponding to a 1.49--1.58$\times$ speedup. The improvement is consistent in both execution directions: forward execution improves by 1.60--1.68$\times$, while backward execution improves by 1.44--1.53$\times$.


The balanced-routing result shows that the optimized taskflow remains effective as the EP group grows. The baseline total latency increases from 44.2 ms at EP4 to 48.9 ms at EP16, reflecting the growing cost of the conventional full-device, collective-communication path. In contrast, \sys\ stays within a narrow 29.6--31.1 ms range because communication, vector work, and GMM computation are exposed as schedulable tasks inside the unified AIC/AIV runtime. These numbers evaluate the full optimized configuration described in Section~\ref{sec:eval_implementation}; component-level attribution is outside this main-results comparison.

\begin{figure}[t]
  \centering
  \includegraphics[width=0.8\linewidth]{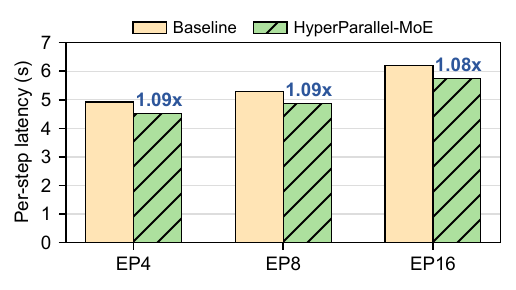}
  \caption{End-to-end latency for one training step with sampled natural routing. Bar annotations report total step-level speedup over the standard operator-by-operator baseline.}
  \label{fig:natural_latency}
\end{figure}
\para{End-to-end training step.} Figure~\ref{fig:natural_latency} reports the latency of one complete training step with sampled natural routing. Unlike the balanced module-level setting, this measurement includes the full training loop around the transformed MoE-FFN path, including unchanged attention and dense-model computation, framework overhead, DVM-level optimizations in the baseline, and naturally produced expert token counts. We use this result to quantify how much of the module-level MoE-FFN improvement remains visible after it is amortized across the complete training step.

For the measured step, \sys\ achieves 1.08--1.09$\times$ end-to-end speedups across EP4, EP8, and EP16. These gains are smaller than the Dispatch-to-Combine speedups because MoE-FFN is only one part of the training critical path: unchanged model computation and framework execution still dominate a large fraction of the step. The step-level result therefore should be interpreted as the diluted effect of a faster MoE-FFN path inside a complete training workflow, rather than as a module-only speedup.

\FloatBarrier

\section{Microbenchmarks}
\label{sect:microbench}

Section~\ref{sect:eval} reports both Dispatch-to-Combine MoE-FFN module latency and end-to-end training-step latency after communication, computation, synchronization, and ordering optimizations are applied together. This section complements that evaluation with focused microbenchmarks that isolate two design assumptions behind \sys's tile-level scheduler. First, decomposing an operator chain into interleaved tiles should expose useful L2 cache reuse of producer outputs without being offset by tile-management overhead. Second, once a workload is taskized, scheduling decisions should be compiled into SSC rather than made by an online dynamic scheduler on the device critical path.

We use a SwiGLU\,+\,Add operator pair as the workload. The pair is small enough to expose scheduling and cache effects clearly, but it has the same producer-consumer structure as the vector-side stages in MoE-FFN: SwiGLU produces an intermediate tensor that is immediately consumed by a downstream elementwise operator. All measurements are collected on Ascend A3. The row dimension $M$ is varied while the SwiGLU input shape is $M\times4096$ and the Add input shape is $M\times2048$.

\subsection{Tile Interleaving and L2 Cache Reuse}
\label{sec:microbench_l2}

This experiment compares the conventional serial execution of SwiGLU followed by Add with a statically scheduled tile-interleaved execution. In the serial mode, each operator runs as a complete kernel-sized stage: SwiGLU finishes all rows before Add begins. In the interleaved mode, both operators are split along the row dimension, and the static schedule places each Add tile soon after the corresponding SwiGLU tile. This is a controlled instance of the mechanism described in Sections~\ref{sec:design_runtime} and~\ref{sec:design_order}: the unified launch lets adjacent tasks pass intermediate data through L2 cache, and the fixed schedule can place related tasks close enough to increase L2 reuse.

\begin{figure}[t]
  \centering
  \begin{subfigure}[t]{0.49\columnwidth}
    \centering
    \includegraphics[width=\linewidth]{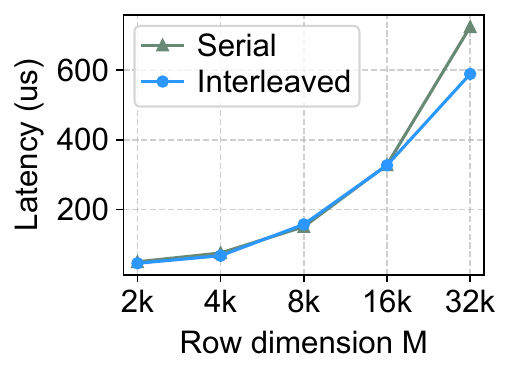}
    \caption{Execution latency}
    \label{fig:swiglu_add_execution_time}
  \end{subfigure}\hfill
  \begin{subfigure}[t]{0.49\columnwidth}
    \centering
    \includegraphics[width=\linewidth]{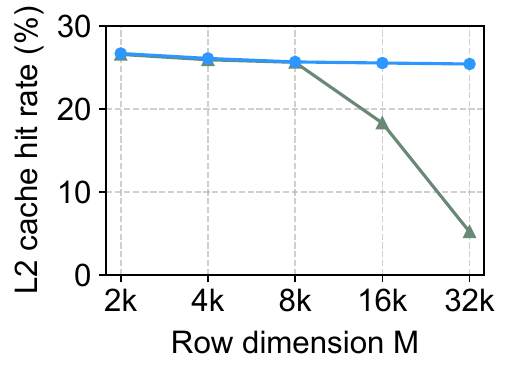}
    \caption{L2 cache hit rate}
    \label{fig:swiglu_add_l2_hit_rate}
  \end{subfigure}
  \caption{SwiGLU+Add cache microbenchmarks under serial and tile-interleaved execution. Left: execution latency. Right: L2 cache hit rate.}
  \label{fig:swiglu_add_microbench}
\end{figure}

Figure~\ref{fig:swiglu_add_execution_time} shows that tile interleaving becomes most valuable when the working set is large enough for the serial producer-consumer distance to exceed useful cache residency. At $M=32$K, tile interleaving reduces latency from 723.29\,$\mu$s to 588.38\,$\mu$s, corresponding to a 1.23$\times$ speedup. Figure~\ref{fig:swiglu_add_l2_hit_rate} explains this latency reduction: the L2 cache hit rate increases from 5.20\% to 25.44\%, reaching 4.9$\times$ that of the serial path. The higher hit rate lets the Add tile read more SwiGLU intermediates from L2 cache instead of HBM; on Ascend A3, L2 cache read bandwidth is more than 4$\times$ higher than HBM bandwidth, so avoiding HBM round trips directly reduces latency. This is the expected effect of the static order: after a SwiGLU tile produces its output, the matching Add tile reads the intermediate before it is displaced by later rows.

For smaller $M$, both execution modes already fit more comfortably within the cache behavior of the device, so the latency gap is small. The 8K point shows a minor slowdown from interleaving, which is consistent with a regime where the extra tile synchronization and queue-consumption work is not yet compensated by additional locality. The important observation is that the interleaved schedule does not rely on changing the SwiGLU or Add implementation. The benefit comes from exposing tile boundaries and choosing a cache-friendly legal order, which is the same scheduling lever used by \sys\ for larger MoE-FFN taskflows.

\subsection{Static vs.\ Dynamic Scheduling Overhead}
\label{sec:microbench_static_dynamic}

The previous subsection uses the serial operator sequence as the baseline for the benefit of taskization and interleaving. We therefore isolate whether, after the workload has already been decomposed into tiles, the tile order should be decided dynamically at runtime or statically through SSC. Both configurations in this experiment execute the taskized SwiGLU\,+\,Add workload with the same event dependencies. The dynamic path performs runtime task dispatch and scheduling decisions on the device path, while the static path consumes the precompiled CTQ/VTQ order and event metadata from SSC.

\begin{figure}[t]
  \centering
  \includegraphics[width=0.8\columnwidth]{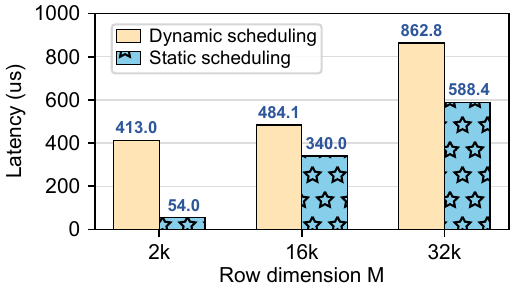}
  \caption{SwiGLU+Add latency under dynamic and static scheduling.}
  \label{fig:swiglu_add_schedule_latency}
\end{figure}

The scheduling-overhead measurements show that dynamic scheduling is too expensive for the tile sizes targeted by \sys. The measured dynamic task dispatch cost is about 2.36\,$\mu$s per AIV task, while the static path costs about 0.1\,$\mu$s per task. This per-task gap dominates the small-$M$ workload: as shown in Figure~\ref{fig:swiglu_add_schedule_latency}, at $M=2$K, dynamic scheduling takes 413.00\,$\mu$s, whereas static scheduling takes only 54.00\,$\mu$s; the dynamic scheduling time is 7.65$\times$ that of static scheduling. Even at $M=32$K, where computation is large enough to amortize part of the overhead, dynamic scheduling still takes 862.80\,$\mu$s compared with 588.38\,$\mu$s for static scheduling; its execution time is 1.47$\times$ that of the static path.

These results justify the static scheduling choice in Section~\ref{sec:design_schedule}. In the target MoE training setting, operator structure, parallel configuration, and dominant shape buckets are known before steady-state execution. Compiling tile order, dependency thresholds, and event assignments into SSC removes per-tile scheduling decisions from the runtime critical path, while still allowing the scheduler to choose locality-aware and overlap-aware legal task orders offline. Dynamic scheduling remains useful for highly unpredictable shape or dependency patterns. For MoE training, however, the operator structure and dominant shape buckets can usually be compiled ahead of time; in this setting, online flexibility is outweighed by per-task dispatch cost on the device critical path.

\section{Other Related Work}
\label{sect:relatedwork}
In this section, we focus on the closest recent MoE mega-kernel systems. COMET~\cite{comet} and UniEP~\cite{uniep} both target distributed MoE execution on GPU clusters, with a focus on communication-computation fusion and fine-grained overlap for expert parallelism. COMET decomposes MoE execution, analyzes the dependencies between AllToAll communication and expert computation, and reorders tasks to reduce exposed communication latency. UniEP uses a MegaKernel to unify multiple EP optimization paths and relies on parameter search to adapt to different configurations. These systems share with \sys\ the goal of breaking coarse-grained operator boundaries and reducing EP communication and scheduling overhead. The key difference is that \sys\ targets the AIC/AIV heterogeneous execution model of Ascend NPUs~\cite{squeezing,accelerating}. Rather than only optimizing the overlap between EP communication and expert GMM, \sys\ organizes the full MoE-FFN path, including Dispatch/Combine, GMM, SwiGLU, and the corresponding backward operators, as a static heterogeneous taskflow. The benefit of this full-path organization comes not only from reducing kernel launches, host-side synchronization, and intermediate tensor traffic through operator fusion, but also from the cross-operator scheduling space exposed by the unified taskflow: communication, GMM, vector operators, and backward branches can be partitioned, synchronized, and ordered within the same dependency graph, improving AIC/AIV overlap in both forward and backward execution.

MegaMoE~\cite{megamoe}, described in the DeepSeek-V4 technical report~\cite{deepseekv4}, demonstrates the engineering value of production-grade MoE mega-kernels~\cite{deepgemm}. MegaMoE targets NVIDIA SM90/SM100 CUDA platforms and fuses the forward MoE-FFN taskflow into a single mega-kernel to overlap NVLink communication with Tensor Core computation. Similar to MegaMoE, \sys\ transforms MoE execution from a sequence of kernel launches into a fine-grained pipeline within a single fused execution window. However, the two systems differ in their goals and scope. MegaMoE currently mainly serves inference and does not cover MoE backward training operators, whereas \sys\ is designed to cover both forward and backward MoE-FFN taskflows. It supports unified scheduling of Dispatch/Combine, GMM, SwiGLU, and gradient operators for both training and inference, while preserving existing optimized operator implementations where possible.

Overall, \sys\ targets a different hardware setting: these systems primarily optimize MoE execution on GPUs through mega-kernel fusion, whereas \sys\ focuses on static heterogeneous task scheduling for MoE execution on Ascend NPUs.




\section{Conclusion}
\label{sect:conclusion}

This paper presents \sys, a fine-grained scheduling framework for MoE training on Ascend NPUs. Instead of executing MoE-FFN as a sequence of full-device operators, \sys\ compiles it into a static tile-level taskflow that coordinates matrix, vector, and communication work while reusing optimized operator bodies. Integrated into the MindFormers stack, the Ascend A3 prototype reduces Dispatch-to-Combine MoE-FFN latency by 1.49--1.58$\times$ under balanced routing, with a sampled natural-routing window showing 1.08--1.09$\times$ end-to-end training speedup. These results indicate that scheduling-layer optimization can expose AIC/AIV parallelism on existing hardware; full-model evaluation and broader operator coverage remain future work.

\bibliographystyle{ACM-Reference-Format}
\bibliography{reference}

@misc{deepseekmoe,
  author    = {Dai, Damai and Deng, Chengqi and Zhao, Chenggang and Xu, R.X. and Gao, Huazuo and Chen, Deli and Li, Jiashi and Zeng, Wangding and Yu, Xingkai and Wu, Y. and others},
  title     = {{DeepSeekMoE}: Towards Ultimate Expert Specialization in Mixture-of-Experts Language Models},
  year      = {2024},
  eprint    = {2401.06066},
  archivePrefix = {arXiv},
  primaryClass = {cs.CL},
  url       = {https://arxiv.org/abs/2401.06066}
}

@inproceedings{flashattention,
  author    = {Dao, Tri and Fu, Daniel Y. and Ermon, Stefano and Rudra, Atri and R{\'e}, Christopher},
  title     = {{FlashAttention}: Fast and Memory-Efficient Exact Attention with {IO}-Awareness},
  booktitle = {Advances in Neural Information Processing Systems},
  volume    = {35},
  pages     = {16344--16359},
  publisher = {Curran Associates, Inc.},
  address   = {Red Hook, NY, USA},
  year      = {2022},
  url       = {https://arxiv.org/abs/2205.14135}
}

@inproceedings{flashmoe,
 author = {Aimuyo, Osayamen and Oh, Byungsoo and Singh, Rachee},
 address = {Red Hook, NY, USA},
 booktitle = {Advances in Neural Information Processing Systems},
 editor = {D. Belgrave and C. Zhang and H. Lin and R. Pascanu and P. Koniusz and M. Ghassemi and N. Chen},
 pages = {100676--100699},
 publisher = {Curran Associates, Inc.},
 title = {FlashMoE: Fast Distributed MoE in a Single Kernel},
 url = {https://proceedings.neurips.cc/paper_files/paper/2025/file/918d938bd209e5b56072777366f8a211-Paper-Conference.pdf},
 volume = {38},
 year = {2025}
}

@inproceedings {pipethreader,
	author = {Yu Cheng and Lei Wang and Yining Shi and Yuqing Xia and Lingxiao Ma and Jilong Xue and Yang Wang and Zhiwen Mo and Feiyang Chen and Fan Yang and Mao Yang and Zhi Yang},
	title = {{PipeThreader}: {Software-Defined} Pipelining for Efficient {DNN} Execution},
	booktitle = {19th USENIX Symposium on Operating Systems Design and Implementation (OSDI 25)},
	year = {2025},
	isbn = {978-1-939133-47-2},
	address = {Boston, MA},
	pages = {767--783},
	url = {https://www.usenix.org/conference/osdi25/presentation/cheng},
	publisher = {USENIX Association},
	month = jul
}

@misc{tritondistributed,
      title={Triton-distributed: Programming Overlapping Kernels on Distributed AI Systems with the Triton Compiler}, 
      author={Size Zheng and Wenlei Bao and Qi Hou and Xuegui Zheng and Jin Fang and Chenhui Huang and Tianqi Li and Haojie Duanmu and Renze Chen and Ruifan Xu and Yifan Guo and Ningxin Zheng and Ziheng Jiang and Xinyi Di and Dongyang Wang and Jianxi Ye and Haibin Lin and Li-Wen Chang and Liqiang Lu and Yun Liang and Jidong Zhai and Xin Liu},
      year={2025},
      eprint={2504.19442},
      archivePrefix={arXiv},
      primaryClass={cs.DC},
      url={https://arxiv.org/abs/2504.19442}, 
}

@inproceedings{pipedream,
author = {Narayanan, Deepak and Harlap, Aaron and Phanishayee, Amar and Seshadri, Vivek and Devanur, Nikhil and Granger, Greg and Gibbons, Phil and Zaharia, Matei},
title = {PipeDream: Generalized Pipeline Parallelism for DNN Training},
booktitle = {Proceedings of the 27th ACM Symposium on Operating Systems Principles},
year = {2019},
month = {October},
address = {Huntsville, ON, Canada},
publisher = {Association for Computing Machinery},
numpages = {15},
doi = {10.1145/3341301.3359646},
url = {https://www.microsoft.com/en-us/research/publication/pipedream-generalized-pipeline-parallelism-for-dnn-training/},
}

@misc{megatronlm,
      title={Megatron-LM: Training Multi-Billion Parameter Language Models Using Model Parallelism}, 
      author={Mohammad Shoeybi and Mostofa Patwary and Raul Puri and Patrick LeGresley and Jared Casper and Bryan Catanzaro},
      year={2019},
      eprint={1909.08053},
      archivePrefix={arXiv},
      primaryClass={cs.CL},
      url={https://arxiv.org/abs/1909.08053}, 
}

@misc{flux,
      title={FLUX: Fast Software-based Communication Overlap On GPUs Through Kernel Fusion}, 
      author={Li-Wen Chang and Wenlei Bao and Qi Hou and Chengquan Jiang and Ningxin Zheng and Yinmin Zhong and Xuanrun Zhang and Zuquan Song and Chengji Yao and Ziheng Jiang and Haibin Lin and Xin Jin and Xin Liu},
      year={2024},
      eprint={2406.06858},
      archivePrefix={arXiv},
      primaryClass={cs.LG},
      url={https://arxiv.org/abs/2406.06858}, 
}

@inproceedings{comet,
 author = {Zhang, Shulai and Zheng, Ningxin and Lin, Haibin and Jiang, Ziheng and Bao, Wenlei and Jiang, Chengquan and Hou, Qi and Cui, Weihao and Zheng, Size and Chang, Li-Wen and Chen, Quan and Liu, Xin},
 address = {Santa Clara, CA, USA},
 booktitle = {Proceedings of Machine Learning and Systems},
 editor = {M. Zaharia and G. Joshi and Y. Lin},
 numpages = {16},
 publisher = {MLSys},
 title = {COMET: Fine-grained Computation-communication Overlapping for Mixture-of-Experts},
 url = {https://proceedings.mlsys.org/paper_files/paper/2025/file/e27ea0cd50b798ff8942caf9203f0992-Paper-Conference.pdf},
 volume = {7},
 year = {2025}
}

@inproceedings{lancet,
  author    = {Chenyu Jiang and Ye Tian and Zhen Jia and Shuai Zheng and Chuan Wu and Yida Wang},
  title     = {Lancet: Accelerating Mixture-of-Experts Training via Whole Graph Computation-Communication Overlapping},
  booktitle = {Proceedings of Machine Learning and Systems},
  volume    = {6},
  publisher = {MLSys},
  address   = {Santa Clara, CA, USA},
  numpages  = {13},
  year      = {2024},
  eprint    = {2404.19429},
  archivePrefix = {arXiv},
  primaryClass = {cs.DC},
  url       = {https://proceedings.mlsys.org/paper_files/paper/2024/file/339caf45a6fa281cae8adc6465343464-Paper-Conference.pdf}
}

@misc{deepep,
  title={DeepEP},
  author={{DeepSeek-AI}},
  year={2025},
  howpublished={\url{https://github.com/deepseek-ai/DeepEP}}
}

@inproceedings{megascale,
  author    = {Jiang, Ziheng and Lin, Haibin and Zhong, Yinmin and Huang, Qi and Chen, Yangrui and Zhang, Zhi and Peng, Yanghua and Li, Xiang and Xie, Cong and Nong, Shibiao and Jia, Yulu and He, Sun and Chen, Hongmin and Bai, Zhihao and Hou, Qi and Yan, Shipeng and Zhou, Ding and Sheng, Yiyao and Jiang, Zhuo and Xu, Haohan and Wei, Haoran and Zhang, Zhang and Nie, Pengfei and Zou, Leqi and Zhao, Sida and Xiang, Liang and Liu, Zherui and Li, Zhe and Jia, Xiaoying and Ye, Jianxi and Jin, Xin and Liu, Xin},
  title     = {{MegaScale}: Scaling {LLM} Training to More Than 10,000 {GPUs}},
  booktitle = {21st USENIX Symposium on Networked Systems Design and Implementation (NSDI 24)},
  year      = {2024},
  isbn      = {978-1-939133-39-7},
  address   = {Santa Clara, CA},
  pages     = {745--760},
  url       = {https://www.usenix.org/conference/nsdi24/presentation/jiang-ziheng},
  publisher = {USENIX Association}
}

@misc{enec,
      title={ENEC: A Lossless AI Model Compression Method Enabling Fast Inference on Ascend NPUs}, 
      author={Jinwu Yang and Jiaan Wu and Zedong Liu and Xinyang Ma and Hairui Zhao and Yida Gu and Yuanhong Huang and Xingchen Liu and Wenjing Huang and Zheng Wei and Jing Xing and Yili Ma and Qingyi Zhang and Baoyi An and Zhongzhe Hu and Shaoteng Liu and Xia Zhu and Jiaxun Lu and Guangming Tan and Dingwen Tao},
      year={2026},
      eprint={2604.03298},
      archivePrefix={arXiv},
      primaryClass={cs.AR},
      url={https://arxiv.org/abs/2604.03298}, 
}

@inproceedings{squeezing,
author = {Zhou, Yuhang and Wang, Zhibin and Liu, Guyue and Li, Shipeng and Lin, Xi and Wang, Zibo and Wang, Yongzhong and Wei, Fuchun and Zhang, Jingyi and Hu, Zhiheng and Liu, Yanlin and Li, Chunsheng and Zhang, Ziyang and Wang, Yaoyuan and Zhou, Bin and Dou, Wanchun and Chen, Guihai and Tian, Chen},
title = {Squeezing Operator Performance Potential for the Ascend Architecture},
year = {2025},
isbn = {9798400710797},
publisher = {Association for Computing Machinery},
address = {New York, NY, USA},
url = {https://doi.org/10.1145/3676641.3716243},
doi = {10.1145/3676641.3716243},
booktitle = {Proceedings of the 30th ACM International Conference on Architectural Support for Programming Languages and Operating Systems, Volume 2},
pages = {1156–1171},
numpages = {16},
location = {Rotterdam, Netherlands},
series = {ASPLOS '25}
}

@inproceedings {accelerating,
	author = {Yuhang Zhou and Zibo Wang and Zhibin Wang and Ruyi Zhang and Chen Tian and Xiaoliang Wang and Wanchun Dou and Guihai Chen and Bingqiang Wang and Yonghong Tian and Yan Zhang and Hui Wang and Fuchun Wei and Boquan Sun and Jingyi Zhang and Bin She and Teng Su and Yifan Yao and Chunsheng Li and Ziyang Zhang and Yaoyuan Wang and Bin Zhou and Guyue Liu},
	title = {Accelerating Model Training on Ascend Chips: An Industrial System for Profiling, Analysis and Optimization},
	booktitle = {2025 USENIX Annual Technical Conference (USENIX ATC 25)},
	year = {2025},
	isbn = {978-1-939133-48-9},
	address = {Boston, MA},
	pages = {1387--1408},
	url = {https://www.usenix.org/conference/atc25/presentation/zhou},
	publisher = {USENIX Association},
	month = jul
}

@inproceedings {deepserve,
	author = {Junhao Hu and Jiang Xu and Zhixia Liu and Yulong He and Yuetao Chen and Hao Xu and Jiang Liu and Jie Meng and Baoquan Zhang and Shining Wan and Gengyuan Dan and Zhiyu Dong and Zhihao Ren and Changhong Liu and Tao Xie and Dayun Lin and Qin Zhang and Yue Yu and Hao Feng and Xusheng Chen and Yizhou Shan},
	title = {{DEEPSERVE}: Serverless Large Language Model Serving at Scale},
	booktitle = {2025 USENIX Annual Technical Conference (USENIX ATC 25)},
	year = {2025},
	isbn = {978-1-939133-48-9},
	address = {Boston, MA},
	pages = {57--72},
	url = {https://www.usenix.org/conference/atc25/presentation/hu-junhao},
	publisher = {USENIX Association},
	month = jul
}

@misc{deepseekv2,
      title={DeepSeek-V2: A Strong, Economical, and Efficient Mixture-of-Experts Language Model}, 
      author={{DeepSeek-AI}},
      year={2024},
      eprint={2405.04434},
      archivePrefix={arXiv},
      primaryClass={cs.CL},
      url={https://arxiv.org/abs/2405.04434}, 
}

@misc{deepseekv3,
      title={DeepSeek-V3 Technical Report}, 
      author={{DeepSeek-AI}},
      year={2024},
      eprint={2412.19437},
      archivePrefix={arXiv},
      primaryClass={cs.CL},
      url={https://arxiv.org/abs/2412.19437}, 
}

@misc{mixtral,
      title={Mixtral of Experts}, 
      author={Albert Q. Jiang and Alexandre Sablayrolles and Antoine Roux and Arthur Mensch and Blanche Savary and Chris Bamford and Devendra Singh Chaplot and Diego de las Casas and Emma Bou Hanna and Florian Bressand and Gianna Lengyel and Guillaume Bour and Guillaume Lample and Lélio Renard Lavaud and Lucile Saulnier and Marie-Anne Lachaux and Pierre Stock and Sandeep Subramanian and Sophia Yang and Szymon Antoniak and Teven Le Scao and Théophile Gervet and Thibaut Lavril and Thomas Wang and Timothée Lacroix and William El Sayed},
      year={2024},
      eprint={2401.04088},
      archivePrefix={arXiv},
      primaryClass={cs.LG},
      url={https://arxiv.org/abs/2401.04088}, 
}

@misc{qwen25,
      title={Qwen2.5 Technical Report}, 
      author={{Qwen Team}},
      year={2024},
      eprint={2412.15115},
      archivePrefix={arXiv},
      primaryClass={cs.CL},
      url={https://arxiv.org/abs/2412.15115}, 
}

@misc{mindspore,
  author       = {{MindSpore Contributors}},
  title        = {{MindSpore}},
  howpublished = {\url{https://www.mindspore.cn/}},
  year         = {2020}
}

@misc{mindformers,
  author       = {{MindSpore Contributors}},
  title        = {{MindSpore Transformers}},
  howpublished = {\url{https://www.mindspore.cn/mindformers/docs/en/master/mindformers.html}},
  year         = {2024}
}

@inproceedings{outrageous,
  author       = {Noam Shazeer and
                  Azalia Mirhoseini and
                  Krzysztof Maziarz and
                  Andy Davis and
                  Quoc V. Le and
                  Geoffrey E. Hinton and
                  Jeff Dean},
  title        = {Outrageously Large Neural Networks: The Sparsely-Gated Mixture-of-Experts
                  Layer},
  address      = {Toulon, France},
  booktitle    = {5th International Conference on Learning Representations, {ICLR} 2017,
                  Toulon, France, April 24-26, 2017, Conference Track Proceedings},
  numpages     = {17},
  publisher    = {OpenReview.net},
  year         = {2017},
  url          = {https://openreview.net/forum?id=B1ckMDqlg},
  bibsource    = {dblp computer science bibliography, https://dblp.org}
}

@article{switch,
  author  = {William Fedus and Barret Zoph and Noam Shazeer},
  title   = {Switch Transformers: Scaling to Trillion Parameter Models with Simple and Efficient Sparsity},
  journal = {Journal of Machine Learning Research},
  year    = {2022},
  volume  = {23},
  number  = {120},
  pages   = {1--39},
  url     = {http://jmlr.org/papers/v23/21-0998.html}
}

@misc{uniep,
      title={UniEP: Unified Expert-Parallel MoE MegaKernel for LLM Training}, 
      author={Size Zheng and Xuegui Zheng and {Li-wen} Chang and Jidong Zhai},
      year={2026},
      eprint={2604.19241},
      archivePrefix={arXiv},
      primaryClass={cs.DC},
      url={https://arxiv.org/abs/2604.19241}, 
}

@misc{deepgemm,
      title={DeepGEMM: clean and efficient BLAS kernel library on GPU}, 
      author={Chenggang Zhao and Zhean Xu and Liang Zhao and Jiashi Li and Chenhao Xu and Anyi Xu and Shengyu Liu and Kexing Zhou and Kuai Yu},
      year={2025},
      publisher = {GitHub},
      howpublished = {\url{https://github.com/deepseek-ai/DeepGEMM}},
}

@misc{deepseekv4,
      author={{DeepSeek-AI}},
      title={{DeepSeek-V4}: Towards Highly Efficient Million-Token Context Intelligence},
      year={2026},
      howpublished={Technical report. \url{https://huggingface.co/deepseek-ai/DeepSeek-V4-Pro/blob/main/DeepSeek_V4.pdf}},
      note={Accessed May 19, 2026}
}

@misc{megamoe,
  title        = {MegaMoE},
  author       = {{DeepSeek-AI}},
  year         = {2026},
  howpublished = {\url{https://github.com/deepseek-ai/DeepGEMM/pull/304}},
  note         = {Merged Apr. 17, 2026}
}

@INPROCEEDINGS{dhellam,
  author={Wang, Haiquan and Ruan, Chaoyi and He, Jia and Ruan, Jiaqi and Tang, Chengjie and Ma, Xiaosong and Li, Cheng},
  booktitle={2025 IEEE 43rd International Conference on Computer Design (ICCD)}, 
  title={DHeLlam: General-Purpose, Automatic Micro-Batch Co-Execution for Distributed LLM Training}, 
  year={2025},
  volume={},
  number={},
  pages={70-78},
  doi={10.1109/ICCD65941.2025.00017}}

\end{document}